%% file: main.tex
\documentclass[sigconf]{acmart}

\usepackage[T1]{fontenc}
\usepackage{xcolor}
\usepackage{listings}
\usepackage{booktabs}
\usepackage{xspace}
\usepackage{pifont}
\usepackage{xurl}
\usepackage{comment}
\usepackage{soul}
\usepackage{subcaption}
\usepackage[inline]{enumitem}
\usepackage{mathtools}
\usepackage{hyperref}
\usepackage{cleveref}
\usepackage{algpseudocode,algorithm}
\usepackage{tikz}
\usepackage{multirow}
\usepackage{makecell}
\usepackage{etoolbox}
\usepackage{listings}
\usepackage{booktabs}

\usepackage{multirow}
\usepackage{array}
\usepackage{enumitem}

\newcolumntype{P}[1]{>{\centering\arraybackslash}p{#1}}
\usepackage{pifont}

\renewcommand\footnotetextcopyrightpermission[1]{} %
\definecolor{ForestGreen}{RGB}{34,139,34}
\definecolor{BrickRed}{RGB}{203,65,84}
\definecolor{bblue}{RGB}{30, 90, 170}
\definecolor{softorange}{RGB}{230, 126, 34}
\definecolor{freshgreen}{RGB}{46, 160, 67}

\newcommand\Smaller{\fontsize{6.75}{8.5}\selectfont}

\lstdefinestyle{cyaml}{
	aboveskip=0pt,
	belowskip=0pt,
	keywordstyle=\color{bblue},
	keywords={FROM,COPY,RUN,WORKDIR,ENV},
	classoffset=1,
	morekeywords={8080,apiVersion,kind,metadata,namespace,probe,selector,exclude,include,spec},
}

\begin{document}

\date{}

\title{ORCA: Unveiling Obscure Containers In The Wild}

\author{Jacopo Bufalino}
\affiliation{%
  \institution{Cnam, Cedric}
  \city{Paris}
  \country{France}
}
\additionalaffiliation{%
  \institution{Aalto University}
  \city{Espoo}
  \country{Finland}
}
\email{jacopo.bufalino@lecnam.net}

\author{Agathe Blaise}
\affiliation{%
  \institution{Thales SIX GTS France}
  \city{Gennevilliers}
  \country{France}
}
\email{agathe.blaise@thalesgroup.com}

\author{Stefano Secci}
\affiliation{%
  \institution{Cnam, Cedric}
  \city{Paris}
  \country{France}
}
\email{stefano.secci@cnam.fr}

\begin{abstract}
Modern software development increasingly depends on open-source libraries and third-party components, which are often encapsulated into containerized environments. While improving the development and deployment of applications, this approach introduces security risks, particularly when outdated or vulnerable components are inadvertently included in production environments. Software Composition Analysis (SCA) is a critical process that helps identify and manage packages and dependencies inside a container. However, unintentional modifications to the container filesystem can lead to incomplete container images, which compromise the reliability of SCA tools. In this paper, we examine the limitations of both cloud-based and open-source SCA tools when faced with such obscure images. An analysis of 600 popular containers revealed that obscure containers exist in well-known registries and trusted images and that many tools fail to analyze such containers. To mitigate these issues, we propose an obscuration-resilient methodology for container analysis and introduce ORCA (Obscuration-Resilient Container Analyzer), its open-source implementation. We reported our findings to all vendors using their appropriate channels. Our results demonstrate that ORCA effectively detects the content of obscure containers and achieves a median 40\% improvement in file coverage compared to Docker Scout and Syft. 
    \end{abstract}

\maketitle
\pagestyle{plain}

\input{introduction}
\input{background}

\input{methodology}

\input{evaluation}
\input{related}

\input{discussion}

\input{conclusion}

\begin{acks}
We thank our shepherd for valuable guidance and feedback throughout the review process. This work was partly supported by the European Commission under grant n.101120393
(Sec4AI4Sec).
\end{acks}

\bibliographystyle{ACM-Reference-Format}

\bibliography{biblio.bib}
\appendix
    \input{appendix}

\end{document}

%% file: introduction.tex
\section{Introduction}
\label{sec:intro}

Modern software development increasingly relies on open-source libraries and third-party components, which are often encapsulated into containerized environments. Together, software, dependencies, tools, and processes form a complex ecosystem known as the software supply chain. Such networked architecture improves productivity but also increases the attack surface, creating new opportunities for adversaries to infiltrate and compromise software systems. 
Among the different threats to the Software Supply Chain, the issue of vulnerable and outdated components is of primary importance, as acknowledged in the OWASP Top 10 project~\cite{owaspTop10}. The importance of this problem is such that regulators~\cite{sbom,cyberrecilience} have defined policies and mandatory requirements to secure the Software Supply Chain of digital products. In this respect, one of the key documents is the Software Bill of Materials (SBOM), which is a record of the details and relationships of the components included in a digital product. Much like %
labels for physical products, SBOMs can be used to identify outdated or vulnerable components. SBOMs for containers are generated using Software Composition Analysis (SCA)~\cite{imtiaz2021comparative} techniques and tools that automatically analyze the content of the container image to extract package and metadata information. They do that by scanning for known filenames and paths across the different container layers. 

However, such tools work on a best-effort basis and make implicit assumptions about the organization of the container filesystem. In practice, such assumptions do not always hold, as legitimate developers may accidentally modify container contents in ways that hinder vulnerability detection. For instance, this can happen by deleting index files, installing software from source without a package manager, using multi-stage builds, or compressing container layers. These common practices, while not malicious, can hide the presence of vulnerable packages, leading to incomplete or misleading SBOMs. We refer to such as \textit{obscure} containers.

While prior work and industry reports~\cite{ESORICS:2020:Liu,maliciouscompliance} have hinted at the existence of such issues, no study to date has systematically cataloged obscuration techniques, quantified their impact, or provided practical mitigation strategies. This paper is the first, to the best of the authors' knowledge, to formalize container obscuration as a software supply chain vulnerability and analyze its impact.
As such, we establish the following research questions:
\begin{itemize}
    \item \textbf{RQ1}: What are the different types of container obscuration?
    \item \textbf{RQ2}: Are state-of-the-art tools vulnerable to obscuration?
    \item \textbf{RQ3}: Are there obscure containers among popular container registries?
\end{itemize}

We designed a three-stage study to systematically address these research questions. First, we conducted a comprehensive review of prior work on container security to construct a corpus of known and novel obscuration techniques. Second, we applied these techniques to create increasingly obscure containers, which we then evaluated against popular state-of-the-art SCA tools. Finally, we formulated a methodology to identify instances of obscure containers and to assess their prevalence across registries. Building on the findings from the previous stages, we also propose a novel methodology for SCA that is resilient to diverse forms of container obscuration. This methodology was implemented in \emph{ORCA}, an open-source tool that improves file coverage by at least 24\% compared to state-of-the-art alternatives. Our evaluation demonstrates that ORCA can be integrated into CI/CD pipelines, enabling early detection of obscure containers. We disclosed our findings to affected vendors.

The remainder of the paper is organized as follows. We provide the necessary background on containers and container image security in \S\ref{sec:background}. In \S\ref{sec:methodology}, we detail our methodologies: \textit{i}) for analyzing container obscuration, and \textit{ii}) for improving package detection with resilience against obscuration. In \S\ref{sec:evaluation}, we highlight the research questions we address through an extensive analysis of open-source containers. In \S\ref{sec:rw} we analyze the related work on SCA for container images and attacks to the software supply chain. In \S\ref{sec:discussions}, we discuss the key findings, proposed mitigation, and responsible disclosure. Finally, in \S\ref{sec:ccl} we conclude the paper.

%% file: background.tex
\section{Background}
\label{sec:background}

\subsection{Containers: A primer}
A container is a lightweight form of virtualization that allows a collection of processes to run in an isolated user-space environment. Containers consist of an image (the software and libraries) and a set of namespaces that define the running environment. A container runtime provides a user interface and libraries for managing containers. %

A container image represents the filesystem of a container. Due to the isolated nature of containers, the software running within them must include all the needed dependencies and libraries. These dependencies are often reused across multiple containers~\cite{Zhao21}, as are application libraries and packages. Container images are thus composed of layers that, combined, generate the final filesystem. Starting from a root filesystem, each new layer represents a change-set of added, deleted, and modified files. This allows for the storage, caching, and reuse of individual layers, optimizing both space and resource consumption. Deleted or modified files are not removed from the layer; instead, they are marked as whiteouts.
Containers are typically generated from declarative files called Containerfiles, which contain instructions to assemble the final filesystem. A fixed set of instructions is available in~\cite{DockerfileReference}. %

A container image follows a structured format consisting of three key components. First, the layers are provided as a list of archives, each representing a change-set on the filesystem of the image. The manifest describes the image, including references to the layers, their cryptographic digests, and metadata. The configuration defines the order in which layers are applied and includes the commands used to generate each layer. While this configuration reflects how the image is built and run, it is not the same as the original Containerfile. Instead, it is a processed result of the build that includes the actual runtime instructions derived from those build files.

\subsection{Container image security} %

The security of container images is an obvious concern, especially in interconnected cloud computing scenarios. Many studies \cite{NIST:2017:Souppaya,ContainerSecurityIssuesa,ARXIV:2015:Bui,flauzacReviewNativeContainer2020} have meticulously explained and grouped the different threats to containers. These threats can be divided into two broad areas, namely \emph{security of the container execution environment} (e.g., privileges, network, OS isolation, container runtime) and \emph{security of the container image} (e.g., vulnerabilities, exploits, secrets, licenses). The former measures and inspects the behavior of containers by monitoring network connections, processes, and access logs. The latter targets the problem of finding vulnerable files and dependencies in the container filesystem. We focus on this problem, which is typically addressed in two phases: discovery and mapping.

Discovery consists of analysing the image filesystem, typically by listing and analyzing a subset of known paths that are likely to contain package-related information (``paths of interest''). Similarly, not all layers are analyzed, but only the squashed representation of the filesystem (i.e., excluding intermediate changes to the image). This process, known as Software Composition Analysis (SCA), has gained attention in recent years due to the increasing complexity of code dependencies in software and containers. Information about the included packages is consolidated into a machine-readable record known as the SBOM~\cite{sbom}.%
A standard format for SBOM is the System Package Data Exchange (SPDX)~\cite{spdx}, developed by the Linux Foundation.

Mapping involves converting the SBOM into Common Vulnerabilities and Exposures (CVEs) using dedicated databases such as the National Vulnerability Database~\cite{NVD}. Mapping packages to CVEs is not a one-time operation, as new vulnerabilities are discovered every day, making it necessary to execute this task at regular intervals.

%% file: methodology.tex
\section{Methodology}

\begin{figure*}[!htb]
    \centering
    \includegraphics[width=\textwidth,trim=0cm 0.2cm 0cm 0.2cm,clip]{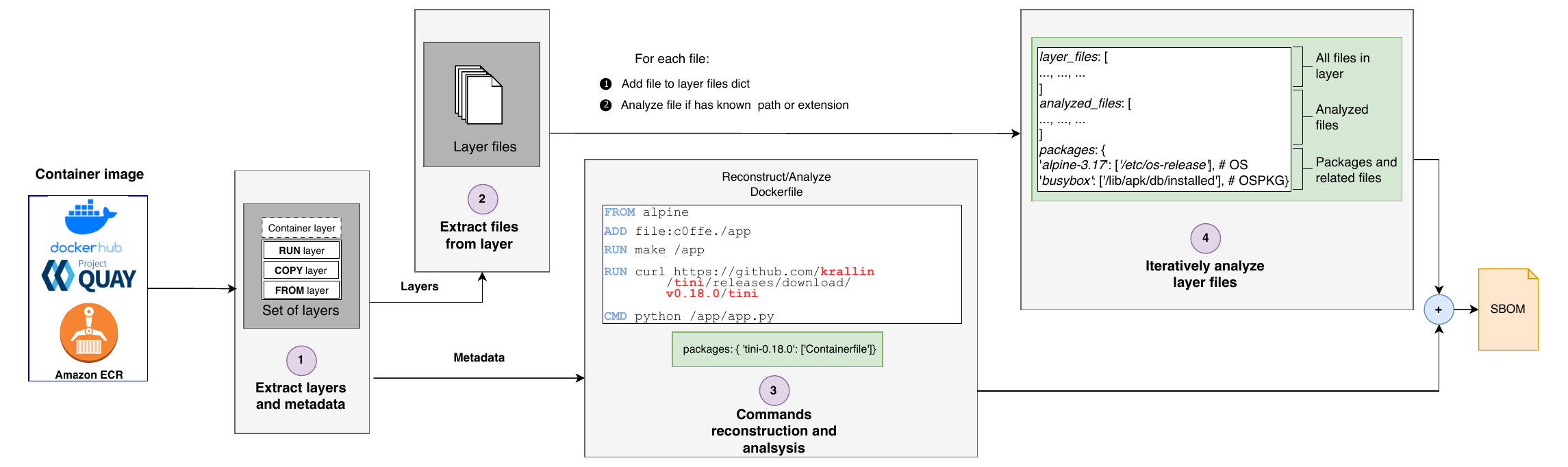}
    \caption{Our layer-by-layer methodology for Software Composition Analysis in containers.}
    \label{fig:sca_container}
\end{figure*}

\label{sec:methodology}
\begin{table*}[!ht]
  \def\arraystretch{1.125}
  \centering
  \begin{tabular}{llllr}
  \toprule
  \textbf{ID} & \textbf{Tactic} & \textbf{Target} & \textbf{Description} & \textbf{Novel} \\
  \cmidrule(lr){1-1} \cmidrule(lr){2-2} \cmidrule(lr){3-3} \cmidrule(lr){4-4} \cmidrule(lr){5-5}
  OS & Operating System Hiding & Operating System files & Modify/Delete OS name and version & \\
  OSPKG & OS Package Manager Hiding & OS Package Manager lockfile & Modify/Delete Package manager lockfile & \\
  DEP & Dependency Hiding & Language lockfile & Modify/Delete programming language lockfile & \\
  COMPRESS & Layer Compression & Every file in the system & Compress the layers of the container & \\
  LINK & File Linking & Any file in the system & Creating links to avoid path scanners & \\
  URL & External sources & Any type of package & Downloading external packages  & \\
  PKG & Package Hiding & Language packages & Modify/Delete installed language dependencies & \checkmark \\
  ALIAS & Alias Creation & Any type of package & Creating alias to avoid path scanners & \checkmark \\
  \bottomrule
  \end{tabular}
  \caption{Summary of obscuration tactics}
  \label{tab:obscuration_tactics}
\end{table*}

In this section, we outline the methodology and assumptions to answer the research questions. 

\subsection{Assumptions and scope}
In this work, we consider the creator of the container image to be a legitimate user who does not intend to harm the system. Therefore, we do not target malicious containers or motivated attackers. We also assume that users do not use binary packers or other mechanisms to purposely hide the content of files or packages. Finally, we assume SCA tools report only the packages and vulnerabilities they find in the image and do not inflate their findings.

\subsection{Systematic review}
To ground our research in plausible obscuration scenarios, we conducted a systematic review of existing work on container image security. This process allowed us to catalog known obscuration tactics and also revealed significant gaps, which led to the identification of previously unknown techniques. For that, we followed the three-step approach as recommended by Petersen\textit{ et al.}~\cite{petersen2015guidelines}.

\paragraph{(1) Source identification}
We gathered information sources from both academia and industry. In detail, we considered research work from the top-tier venues in security and software engineering~\cite{IEEESP:2022:ENCK, IEEESP:2023:Fourne, IEEESP:2023:Melara, 1203227, 264122, duan2020towards,10179304} --- including ACM CCS, NDSS, USENIX Security, IEEE S\&P, ICSE, FSE, ASE --- and published in the last five years. We also searched for papers outside of the top conferences but containing relevant keywords~\cite{CMC:2022:Doan,CCWC:2020:Brady,Zhao21,Javed21}. Finally, we included industry standards~\cite{NIST:2017:Souppaya,NVD_CPE,sbom}, white-papers~\cite{AWSSbom,AzureSBOM,ScanImagesOS,UnderstandingImageLayers0530}, conferences~\cite{maliciouscompliance,google_sbom}, and open-source tools~\cite{clair,grype,cve-bin,Crane,docker_scout,trivy,FunContainerImages}. 

\paragraph{(2) Inclusion and Exclusion Criteria}
Among the initial corpus of sources, we selected the ones discussing attacks on the software supply chain or on container images. In the end, we were left with five academic works~\cite{IEEESP:2023:Melara, 10179304, IEEESP:2022:ENCK, duan2020towards,CCWC:2020:Brady} and two industry ones~\cite{maliciouscompliance,google_sbom}.

\paragraph{(3) Codify results}
Two of the authors independently extracted obscuration tactics and targets from the final artifacts and identified four existing tactics for container image obscuration. %

\paragraph{(4) Novel tactics}
Building upon the systematic review, we identified new obscuration cases that, to the best of the authors' knowledge, have not been previously documented. We found new ways to alter the content of the container images. A summary of the results is available in Table~\ref{tab:obscuration_tactics}. 
\subsection{Obscuration targets and tactics (RQ1)}\label{defining-container-image-obscuration}

In the following, we detail the different container obscuration targets and tactics.

\paragraph{Operating System Hiding} Information about the operating system is stored in specific files under the \emph{/etc} or \emph{/usr} directories. These generally include data about the operating system name (e.g., Ubuntu) and version (e.g., 22.04.4). Security tools use this content to identify OS-specific package versions and vulnerabilities, and if they are edited or removed then those tools will not be able to identify the OS and will report fewer packages and vulnerabilities.

\paragraph{OS Package Manager Hiding} Many operating system libraries and binaries are installed through package managers that depend on Linux distributions (e.g., \emph{apk} for Alpine or \emph{apt} for Debian). When a new package is installed, metadata information such as the generated files, version, and checksum is stored by the manager in a lockfile. Accessing such files is sometimes the only way to find the exact version of a package. %

\paragraph{Dependency Hiding} Similarly to OS packages, programming languages have package managers that track dependencies and their status. Dependency information may be stored in one (e.g., \emph{requirements.txt} in Python) or two files (e.g., \emph{package.json} and \emph{package-lock.json} in JavaScript). In the latter case, one file contains a high-level overview of the dependencies (i.e., in terms of minimum version required or major version used); the other file is more detailed and contains hash values and timestamps. %

\paragraph{Layer Compression} Compressing layers of a container image is a routine and totally legitimate operation. However, compressing layers inevitably removes information about the layers and therefore any potential obscure commands. 

\paragraph{File Linking} Creating links can confuse SCA tools because they usually do not follow links, especially if the link and the linked file belong to different layers.

\paragraph{Package Hiding} The dependencies of a given application can, at times, store information about their version and their dependencies. This is not standardized, so the actual information and the way it is stored may vary. Usually, version information can be found as part of the source code, inside the package license file, or in other files that have the sole purpose of documenting the package itself. Dependency information can be fetched from the software dependency file of the package, and other files may contain hashes of the package's files. 

\paragraph{External sources} Packages downloaded from the Internet are referred to as external software. They manifest either as git repositories or as compiled binaries~\cite{Kawaguchi24}. The danger associated with external sources is that the resulting artifacts (i.e., the binaries) do not leave any provenance information in the container, and therefore SCA tools are unable to identify them.  

\paragraph{Alias Creation} Aliases produce similar effects as links, but they are harder to find because the aliases are usually written in environment variables or files.

\subsection{Detect obscure containers}
\label{subsect:detect_obscuration}

Detecting cases of obscure containers is key to understanding their presence in the wild. For that, we analyze container images layer by layer by extracting them as archives and evaluating their modification history. Together with that, we analyze and parse the metadata information of the container image. We describe hereafter the overall process for detecting obscuration.

\paragraph{(1) Layer and metadata extraction} The first step of our obscuration detection process involves extracting the container metadata (which includes the index file -- see Section~\ref{sec:background}) along with all image layers, which are unpacked as directories.

\paragraph{(2) File extraction} The second step consists of retrieving, for each layer of the image, the list of files within the layer directory. It is important to preserve the ordering of the layers, which we retrieve from the container's metadata.

\paragraph{(3) Containerfile reconstruction and analysis}
The metadata extracted from the container image index file enables the partial reconstruction of the Containerfile content\footnote{\texttt{COPY} and \texttt{ADD} commands will only show the shasum of the files/directories.}.
This reconstruction is essential for detecting instances of \texttt{URL} obscuration. Our approach first interpolates environment variables and arguments, after which pattern matching is applied to individual instructions in order to identify URLs or repositories. The outcome of this procedure is a list of packages.

\paragraph{(4) Obscuration detection}
For each ordered layer, we list the files that have been created, modified, and deleted. This effectively generates a history of the container. We search the history for files with package content using a list of common names and paths (available in Table~\ref{tab:obscuration:patterns} in Appendix~\ref{app:location}). We detect obscuration when the history of such files shows updates or deletions (layer-by-layer analysis of the history), and when we find packages installed from source (using the reconstructed Containerfile). 

\paragraph{False positives}
Our methodology inherently introduces false positives. These stem from the fact that modification or deletion of a package can occur for benign reasons, such as software updates. In such cases, we still include the outdated packages in our results and mark them as obscured. The impact of this approach is mostly visible in the OS package manager dependencies, because application dependencies are rarely updated during the image generation process. We adopt this conservative approach because some software may have been built and statically linked to the old package and may still contain vulnerable code. 

\subsection{Obscuration-Resilient Container Analyzer}

To improve SBOM accuracy, we introduce an open-source methodology named \emph{ORCA} based on layer-by-layer analysis that reconstructs the full modification history of container image layers, drawing inspiration from techniques we previously demonstrated to detect obscuration. This approach increases resilience to obscuration by offering a more detailed and comprehensive analysis. In addition to identifying package metadata files (as other SCA tools do), we also analyze package-related content files, improving both file coverage (the total number of analyzed files) and SBOM completeness. Figure \ref{fig:sca_container} outlines the main steps of our methodology, highlighting key features that enhance obscuration resilience in green.

\paragraph{(1) Layers analysis, (2) File extraction, and (3) Containerfile reconstruction}
The first three steps follow the same process outlined in Section \ref{subsect:detect_obscuration}.

\paragraph{(4) Iterative file analysis} 
The fourth step involves analyzing the files in each layer. For each layer, we record the list of files present, the list of analyzed files, and a dictionary of the packages mapped to their related files, in a \textit{layer analysis report}, updating the SBOM incrementally. We begin by analyzing the OS and its version, and updating the list of analyzed files. Next, we focus on identifying the operating system package manager. Currently, we support the DPKG, RPM, and APK OS package managers (which correspond to the ones we found in our datasets). These package managers have dedicated files or databases with the list of installed files and folders. We use this information to populate the list of analyzed files and the package dictionary. We adopt a similar approach to analyze programming language package managers.

To enhance file coverage, we map files associated with a project according to either the project's structural organization or the installation format imposed by its package manager. For example, in the case of a JavaScript project that relies on \texttt{npm}, all files located in the \texttt{node\_modules} directory can be associated with that project. This strategy ensures that both package-related content and package metadata files are included in the analysis.
While our aim is to maximize file coverage, we avoid redundant analysis and do not individually parse all files, especially those related to previously analyzed files. 

Another unique feature of ORCA is path analysis, which identifies files and dependencies based on their path. This is useful, for instance, when Python packages are installed in non-default locations.

With respect to binaries, we choose to analyze only Go binaries because they include metadata information about the installed dependencies. This choice is in accordance with state-of-the-art tools. In fact, expression-based search of content is slow and ineffective, as there is no common structure for how versions are stored in a binary. Sometimes they may be hard-coded strings, other times they may be fetched from a shared library or computed at runtime by concatenating multiple variables. We refer the reader to existing tools such as the Intel \emph{cve-bin} \cite{cve-bin} for that purpose.

Finally, the packages extracted from the filesystem are added to the ones computed from the Containerfile. The final corpus of packages is then consolidated in an SBOM. 

In summary, our tool \emph{ORCA} benefits from these four key features -- layer-by-layer analysis, increased file coverage, Containerfile analysis, and file path analysis -- to increase its resilience against obscure containers.

%% file: evaluation.tex
\section{Evaluation}
\label{sec:evaluation}

This section starts by describing the dataset and the setup of our experiments. It then introduces the experiments and our answers to the research questions. 
\subsection{Datasets and experiment setup}
\label{sec:exp_setup}
We first describe the container analysis tools (both open-source and proprietary) used for comparison and then introduce the datasets employed for the different experiments.

\paragraph{Scanner tools} Several tools aim to identify software components within container images. For our study, we retrieved the most well-known open-source~\cite{o2023impacts} and commercial tools. We excluded tools that do not target containers specifically (e.g., cve-bin-tool~\cite{cve-bin}) or that are outdated (e.g., Tern~\cite{Tern}). The list of the tools we used is provided in Table \ref{tab:scanners}.  All of the tools have the possibility of downloading an SBOM of the container or of the filesystem in SPDX format. Only three container security tools --- namely, Scout, Trivy, and Grype --- can produce SBOMs and find vulnerabilities locally, while the others require a subscription to the respective cloud provider. Cloud tools cannot be configured or fine-tuned. For each of the OSS tools, we ensured they operated under their best possible conditions. Among them, only Grype could be setup to scan every container layer. 

\begin{table}[!htb]
\centering
\small
\begin{tabular}{lrrc}
 \toprule
 \textbf{Tool} & \textbf{Version} &\textbf{Company} & \textbf{OSS} \\
 \cmidrule(lr){1-1} \cmidrule(lr){2-4}
 Grype\footnotemark~\cite{grype, syft} & 0.77.0 &Anchore &  \checkmark\\  
 DockerScout~\cite{docker_scout} & 1.11.0 &Docker & \\ 
 Trivy~\cite{trivy} & 0.50.2 &Trivy &  \checkmark\\ 
 Artifact registry~\cite{GCloud} & unknown &Google &  \\ 
 Defender for Cloud~\cite{AzureSBOM} & 2.2.9 & Microsoft & \checkmark \footnotemark \\
 Inspector~\cite{AWSSbom} & unknown & Amazon  &\\ \bottomrule
\end{tabular}
\caption{Static analysis tools used for Software Composition Analysis.}
\label{tab:scanners}
\end{table}
\footnotetext[4]{Grype is using Syft for generating the SBOM.}
\footnotetext[5]{Only the SBOM generation tool is open-source.}

\begin{table}[h]
    \centering
    \small
    \begin{tabular}{lrl}
     \toprule
     \textbf{Dataset} & \textbf{Size} & \textbf{Description} \\
     \cmidrule(lr){1-1} \cmidrule(lr){2-3}
     DockerHub Official & 100  & \begin{tabular}{@{}l@{}}Curated images built in \\ collaboration with software \\ maintainers\end{tabular} \\
     DockerHub Bitnami & 100  & \begin{tabular}{@{}l@{}}Images maintained by Bitnami\end{tabular}  \\
     DockerHub Verified & 100  & \begin{tabular}{@{}l@{}}Images from trusted software \\ publishers\end{tabular}\\
     DockerHub OSS & 100  & \begin{tabular}{@{}l@{}}Images published and \\ maintained by open-source \\ projects sponsored by Docker\end{tabular}\\
     Quay.io & 100  & \begin{tabular}{@{}l@{}}Most used container images in \\ Quay.io\end{tabular}\\
     ECR & 100  & \begin{tabular}{@{}l@{}}Images published and \\ maintained by Amazon \end{tabular}\\\bottomrule 
    \end{tabular}
    \caption{Considered datasets and their size (number of containers).}
    \label{tab:datasets}
\end{table}

\paragraph{Obscure container images dataset}
We built our own set of obscure containers to measure the resilience of SBOM tools. We started from the official, non-obscure, \texttt{python:3.10.0} container image from DockerHub that we used as a base to install a simple web server. We purposely selected this popular container as it is based on Debian and Python, both of which are supported by all the container analysis tools. Furthermore, the 3.10 version has many known vulnerabilities, making it a relevant candidate for our obscuration tests.
This base image allowed us to systematically apply a variety of obscuration techniques, targeting different layers of the container, including the OS, OS packages, and language-specific dependencies. We then generated multiple variations of this image, each implementing one or more obscuration techniques.

\paragraph{Container images dataset} 
We assembled a dataset of 600 publicly available containers from various registries, including DockerHub \cite{DockerHub}, Quay.io \cite{QuayIO}, and Amazon Elastic Container Registry (ECR) \cite{ECR}. The online repositories,  maintained by Docker, RedHat, and Amazon Web Services (AWS), respectively, allow developers and organizations to upload, share, and download container images. We selected the 100 most popular container images from DockerHub Official, DockerHub Bitnami, DockerHub Verified, DockerHub OSS, Quay.io, and ECR. Table \ref{tab:datasets} provides a detailed breakdown of the datasets included in this study; they were selected to provide a diverse and representative collection of container images from different maintainers and platforms. This dataset is used to study the prevalence of obscuration in containers.

\subsection{Resilience to obscure images (RQ2)} %

\begin{table*}[!ht]
    \centering
    \small
\begin{tabular}{lrrrrrrrrrrrrrrrr}
\toprule
 &  \multicolumn{2}{c}{\textbf{Trivy}}  &   \multicolumn{2}{c}{\textbf{Syft}}  &  \multicolumn{2}{c}{\textbf{Syft (All)}}  &  \multicolumn{2}{c}{\textbf{Scout}} &  \multicolumn{2}{c}{\textbf{Microsoft}}  &    \multicolumn{2}{c}{\textbf{Gcloud}}  & \multicolumn{2}{c}{\textbf{Amazon}}  & \multicolumn{2}{c}{\textbf{ORCA}}      \\
 \textbf{Technique(s)} & V & P & V & P& V & P& V & P& V & P & V& P& V & P& V & P \\
 \cmidrule(lr){1-1}\cmidrule(lr){2-3} \cmidrule(lr){4-5} \cmidrule(lr){6-7} \cmidrule(lr){8-9} \cmidrule(lr){10-11} \cmidrule(lr){12-13} \cmidrule(lr){14-15} \cmidrule(lr){16-17} 
 BASE (no obscuration)                                     & \color{black}{1164} & \color{black}{441} & \color{black}{625} & \color{black}{448} & \color{black}{625} & \color{black}{448} & \color{black}{123} & \color{black}{585} & \color{black}{154} & \color{black}{429} & \color{black}{722} & \color{black}{441} & \color{black}{471} & \color{black}{587} & \color{black}{2355} & \color{black}{1046} \\\cmidrule(lr){1-17}
 OSPKG                                     & \color{BrickRed}{6}      & \color{BrickRed}{11}    & \color{BrickRed}{25}    & \color{BrickRed}{25}    &    \color{ForestGreen}{625}    &    \color{ForestGreen}{448} & \color{BrickRed}{10}    & \color{BrickRed}{23}    & \color{ForestGreen}{154} & \color{ForestGreen}{429} & \color{BrickRed}{6}     & \color{BrickRed}{12}    & \color{BrickRed}{N/A}   & \color{BrickRed}{0}     & \color{ForestGreen}{2355} & \color{ForestGreen}{1046} \\
 URL                                      & \color{BrickRed}{1164} & \color{BrickRed}{441} & \color{BrickRed}{625} & \color{BrickRed}{448} &    \color{BrickRed}{625}    &    \color{BrickRed}{448} & \color{BrickRed}{123} & \color{BrickRed}{585} & \color{BrickRed}{154} & \color{BrickRed}{429} & \color{BrickRed}{722} & \color{BrickRed}{441} & \color{BrickRed}{471} & \color{BrickRed}{587} & \color{ForestGreen}{2357} & \color{ForestGreen}{1047} \\
 LINK                                      & \color{ForestGreen}{1164} & \color{ForestGreen}{441} & \color{ForestGreen}{625} & \color{ForestGreen}{448} &    \color{ForestGreen}{625}    &    \color{ForestGreen}{448} & \color{ForestGreen}{123} & \color{ForestGreen}{585} & \color{ForestGreen}{154} & \color{ForestGreen}{429} & \color{ForestGreen}{722} & \color{ForestGreen}{441} & \color{ForestGreen}{471} & \color{ForestGreen}{587} & \color{ForestGreen}{2355} & \color{ForestGreen}{1046} \\
 DEP                                       & \color{ForestGreen}{1164} & \color{ForestGreen}{441} & \color{ForestGreen}{625} & \color{ForestGreen}{448} &    \color{ForestGreen}{625}    &    \color{ForestGreen}{448}& \color{ForestGreen}{123} & \color{ForestGreen}{585} & \color{ForestGreen}{154} & \color{ForestGreen}{429} & \color{ForestGreen}{722} & \color{ForestGreen}{441} & \color{ForestGreen}{471} & \color{BrickRed}{579}   & \color{ForestGreen}{2355} & \color{ForestGreen}{1046} \\
 PKG                                       & \color{BrickRed}{1158}   & \color{BrickRed}{430}   & \color{BrickRed}{619}   & \color{BrickRed}{436} &    \color{ForestGreen}{625}    &    \color{ForestGreen}{448}  & \color{BrickRed}{117}   & \color{BrickRed}{573}   & \color{BrickRed}{148}   & \color{ForestGreen}{429} & \color{BrickRed}{716}   & \color{BrickRed}{429}   & \color{BrickRed}{469}   & \color{BrickRed}{576}   & \color{ForestGreen}{2355} & \color{ForestGreen}{1046} \\
 ALIAS                                     & \color{ForestGreen}{1164} & \color{ForestGreen}{441} & \color{ForestGreen}{625} & \color{ForestGreen}{448} &    \color{ForestGreen}{625}    &    \color{ForestGreen}{448} & \color{ForestGreen}{123} & \color{ForestGreen}{585} & \color{ForestGreen}{154} & \color{ForestGreen}{429} & \color{ForestGreen}{722} & \color{ForestGreen}{441} & \color{ForestGreen}{471} & \color{ForestGreen}{587} & \color{ForestGreen}{2355} & \color{ForestGreen}{1046} \\
 COMPRESS                                  & \color{ForestGreen}{1164} & \color{ForestGreen}{441} & \color{ForestGreen}{625} & \color{ForestGreen}{448} &    \color{ForestGreen}{625}    &    \color{ForestGreen}{448}& \color{ForestGreen}{123} & \color{ForestGreen}{585} & \color{ForestGreen}{154} & \color{ForestGreen}{429} & \color{ForestGreen}{722} & \color{ForestGreen}{441} & \color{ForestGreen}{471} & \color{ForestGreen}{587} & \color{ForestGreen}{2355} & \color{ForestGreen}{1046} \\
 OS                                        & \color{ForestGreen}{1164} & \color{ForestGreen}{441} & \color{BrickRed}{9}     & \color{ForestGreen}{448} &    \color{ForestGreen}{625}    &    \color{ForestGreen}{448}& \color{BrickRed}{6}     & \color{BrickRed}{18}    & \color{ForestGreen}{154} & \color{ForestGreen}{429} & \color{BrickRed}{6}     & \color{BrickRed}{12}    & \color{ForestGreen}{471} & \color{ForestGreen}{587} & \color{ForestGreen}{2355} & \color{ForestGreen}{1046} \\\cmidrule(lr){1-17}
 OS\,$+$\,OSPKG                                & \color{BrickRed}{6}      & \color{BrickRed}{11}    & \color{BrickRed}{27}    & \color{BrickRed}{25}    &    \color{ForestGreen}{625}    &    \color{ForestGreen}{448} & \color{BrickRed}{6}     & \color{BrickRed}{23}    & \color{ForestGreen}{154} & \color{ForestGreen}{429} & \color{BrickRed}{6}     & \color{BrickRed}{12}    & \color{BrickRed}{N/A}   & \color{BrickRed}{0}     & \color{ForestGreen}{2355} & \color{ForestGreen}{1046} \\
 DEP\,$+$\,PKG                                 & \color{BrickRed}{1158}   & \color{BrickRed}{430}   & \color{BrickRed}{619}   & \color{BrickRed}{436}   &    \color{ForestGreen}{625}    &    \color{ForestGreen}{448} & \color{BrickRed}{117}   & \color{BrickRed}{573}   & \color{BrickRed}{148}   & \color{ForestGreen}{429} & \color{BrickRed}{716}   & \color{BrickRed}{429}   & \color{BrickRed}{465}   & \color{BrickRed}{568}   & \color{ForestGreen}{2355} & \color{ForestGreen}{1046} \\
 OS\,$+$\,DEP                                  & \color{ForestGreen}{1164} & \color{ForestGreen}{441} & \color{BrickRed}{9}     & \color{ForestGreen}{448} &    \color{ForestGreen}{625}    &    \color{ForestGreen}{448} &\color{BrickRed}{6}     & \color{BrickRed}{18}    & \color{ForestGreen}{154} & \color{ForestGreen}{429} & \color{BrickRed}{6}     & \color{BrickRed}{12}    & \color{ForestGreen}{471} & \color{BrickRed}{579}   & \color{ForestGreen}{2355} & \color{ForestGreen}{1046} \\
 OS\,$+$\,PKG                                  & \color{BrickRed}{1158}   & \color{BrickRed}{430}   & \color{BrickRed}{3}     & \color{BrickRed}{436}   &    \color{ForestGreen}{625}    &    \color{ForestGreen}{448} & \color{BrickRed}{0}     & \color{BrickRed}{6}     & \color{BrickRed}{148}   & \color{ForestGreen}{429} & \color{BrickRed}{N/A}     & \color{BrickRed}{0}     & \color{BrickRed}{469}   & \color{BrickRed}{576}   & \color{ForestGreen}{2355} & \color{ForestGreen}{1046} \\\cmidrule(lr){1-17}
 OS\,$+$\,OSPKG\,$+$\,COMPRESS                     & \color{BrickRed}{6}      & \color{BrickRed}{12}    & \color{BrickRed}{27}    & \color{BrickRed}{25}    &    \color{BrickRed}{27}    &    \color{BrickRed}{25} &  \color{BrickRed}{6}     & \color{BrickRed}{23}    & \color{BrickRed}{0}     & \color{BrickRed}{0}     & \color{BrickRed}{6}     & \color{BrickRed}{12}    & \color{BrickRed}{N/A}   & \color{BrickRed}{0}     & \color{BrickRed}{14}     & \color{BrickRed}{454}   \\
 OS\,$+$\,OSPKG\,$+$\,PKG                          & \color{BrickRed}{0}      & \color{BrickRed}{0}     & \color{BrickRed}{21}    & \color{BrickRed}{13}    &    \color{ForestGreen}{625}    &    \color{ForestGreen}{448} & \color{BrickRed}{0}     & \color{BrickRed}{11}    & \color{BrickRed}{148}   & \color{ForestGreen}{429} & \color{BrickRed}{N/A}     & \color{BrickRed}{0}     & \color{BrickRed}{N/A}   & \color{BrickRed}{0}     & \color{ForestGreen}{2355} & \color{ForestGreen}{1046} \\
 OS\,$+$\,OSPKG\,$+$\,DEP                          & \color{BrickRed}{6}      & \color{BrickRed}{11}    & \color{BrickRed}{27}    & \color{BrickRed}{25}    &    \color{ForestGreen}{625}    &    \color{ForestGreen}{448} & \color{BrickRed}{6}     & \color{BrickRed}{23}    & \color{ForestGreen}{154} & \color{ForestGreen}{429} & \color{BrickRed}{6}     & \color{BrickRed}{12}    & \color{BrickRed}{N/A}   & \color{BrickRed}{0}     & \color{ForestGreen}{2355} & \color{ForestGreen}{1046} \\\cmidrule(lr){1-17}
 OS\,$+$\,OSPKG\,$+$\,DEP\,$+$\,LINK                   & \color{BrickRed}{6}      & \color{BrickRed}{11}    & \color{BrickRed}{27}    & \color{BrickRed}{25}    &    \color{ForestGreen}{625}    &    \color{ForestGreen}{448} & \color{BrickRed}{6}     & \color{BrickRed}{23}    & \color{ForestGreen}{154} & \color{ForestGreen}{429} & \color{BrickRed}{6}     & \color{BrickRed}{12}    & \color{BrickRed}{N/A}   & \color{BrickRed}{0}     & \color{ForestGreen}{2355} & \color{ForestGreen}{1046} \\
 OS\,$+$\,OSPKG\,$+$\,DEP\,$+$\,PKG                    & \color{BrickRed}{0}      & \color{BrickRed}{0}     & \color{BrickRed}{21}    & \color{BrickRed}{13}    &    \color{ForestGreen}{625}    &    \color{ForestGreen}{448} & \color{BrickRed}{0}     & \color{BrickRed}{11}    & \color{BrickRed}{148}   & \color{ForestGreen}{429} & \color{BrickRed}{N/A}     & \color{BrickRed}{0}     & \color{BrickRed}{N/A}   & \color{BrickRed}{0}     & \color{ForestGreen}{2355} & \color{ForestGreen}{1046} \\
 OS\,$+$\,OSPKG\,$+$\,DEP\,$+$\,COMPRESS               & \color{BrickRed}{6}      & \color{BrickRed}{12}    & \color{BrickRed}{27}    & \color{BrickRed}{25}   &    \color{BrickRed}{27}    & \color{BrickRed}{25}  &  \color{BrickRed}{6}     & \color{BrickRed}{23}    & \color{BrickRed}{0}     & \color{BrickRed}{0}     & \color{BrickRed}{6}     & \color{BrickRed}{12}    & \color{BrickRed}{N/A}   & \color{BrickRed}{0}     & \color{BrickRed}{14}     & \color{BrickRed}{454}   \\
 OS\,$+$\,OSPKG\,$+$\,DEP\,$+$\,ALIAS                  & \color{BrickRed}{6}      & \color{BrickRed}{11}    & \color{BrickRed}{25}    & \color{BrickRed}{24}    &    \color{ForestGreen}{625}    &    \color{ForestGreen}{448} & \color{BrickRed}{6}     & \color{BrickRed}{22}    & \color{ForestGreen}{154} & \color{ForestGreen}{429} & \color{BrickRed}{6}     & \color{BrickRed}{12}    & \color{BrickRed}{N/A}   & \color{BrickRed}{0}     & \color{ForestGreen}{2355} & \color{ForestGreen}{1046} \\\cmidrule(lr){1-17}
 OS\,$+$\,OSPKG\,$+$\,DEP\,$+$\,ALIAS\,$+$\,COMPRESS       & \color{BrickRed}{6}      & \color{BrickRed}{12}    & \color{BrickRed}{25}    & \color{BrickRed}{24}    &    \color{BrickRed}{27}    & \color{BrickRed}{25}  &  \color{BrickRed}{6}     & \color{BrickRed}{22}    & \color{BrickRed}{0}     & \color{BrickRed}{0}     & \color{BrickRed}{6}     & \color{BrickRed}{12}    & \color{BrickRed}{N/A}   & \color{BrickRed}{0}     & \color{BrickRed}{14}     & \color{BrickRed}{454}   \\
 OS\,$+$\,OSPKG\,$+$\,DEP\,$+$\,ALIAS\,$+$\,PKG            & \color{BrickRed}{0}      & \color{BrickRed}{0}     & \color{BrickRed}{19}    & \color{BrickRed}{12}    &    \color{ForestGreen}{625}    &    \color{ForestGreen}{448} &\color{BrickRed}{0}     & \color{BrickRed}{10}    & \color{BrickRed}{148}   & \color{ForestGreen}{429} & \color{BrickRed}{N/A}     & \color{BrickRed}{0}     & \color{BrickRed}{N/A}   & \color{BrickRed}{0}     & \color{ForestGreen}{2355} & \color{ForestGreen}{1046} \\\bottomrule
\end{tabular}
    \caption{Comparison between different obscuration types with different tools. The metrics correspond to the number of vulnerabilities and packages identified. Colors indicate obscuration resilience: {\color{BrickRed}{red}} if obscuration hides vulnerabilities or packages from the tool, {\color{ForestGreen}{green}} if obscuration has no impact. N/A indicates that the tool did not scan the container.}
    \label{tab:obscuration_results}
    \end{table*}

This experiment measures the resilience of popular open-source and proprietary SCA tools when analyzing obscure images. For that, we used the container dataset described in Section~\ref{sec:exp_setup}. For each tool and test case, we apply the obscuration technique(s), we generate an SBOM in the SPDX format, and measure the number of detected packages and vulnerabilities~\footnote{The obscured Containerfiles are available at:~\url{https://github.com/kube-security/container-obfuscation-benchmark}}. For Syft, the experiment was conducted both with and without enforcing layer-by-layer analysis to assess its impact on detection accuracy.
We repeated the same experiment using our ORCA methodology~\footnote{ORCA is vailable at: \url{https://github.com/kube-security/orca}. We used Grype to find vulnerabilities on the generated SBOM.}. We define the original, non-obscure container as our baseline.

A tool is considered vulnerable to a technique if the number of vulnerabilities or packages changes with respect to the baseline. Usually, this means that the number of packages or vulnerabilities becomes smaller. However, the \texttt{URL} obscuration is a special case because when we install packages without a package manager, we expect the number of total packages to grow, so if the number stays the same, it means that the tool was unable to recognize such packages.
We selected a subset from all possible permutations of the obscuration techniques. We decided to analyze all individual obscuration techniques and then include combined techniques that obscure different parts of the container image (e.g., OS and package manager). The results are shown in Table~\ref{tab:obscuration_results}.  

\paragraph{Single obscuration}
We started the experiment by analyzing the effect of single instances of obscuration techniques. The \texttt{LINK}, \texttt{ALIAS}, and \texttt{COMPRESS} techniques alone were not effective against any tool. The \texttt{DEP} technique was only effective against Amazon's package count. This test case is interesting because the number of vulnerabilities discovered with dependency obscuration is the same as in the baseline case. This happens because the Python dependencies of the application did not have vulnerabilities at the time of the scan.

Only Syft (with all layers scanning enabled) and Microsoft's tool were not affected by the deletion of the OS package manager file (\texttt{OSPKG}). In the case of Amazon Inspector, if the OS package manager is deleted, the UI does not show any sign of error, but the container is not scanned. 

The table also shows that every tested tool is vulnerable to URL obscuration. This is evident from the fact that tools are unable to find the manually downloaded packages in the container image, and they report the same number of packages and vulnerabilities as the baseline.

Only Syft, in the all-layer configuration, is not affected by \texttt{PKG} obscuration. In the case of Microsoft Defender, the number of detected packages does not change compared to the baseline, but the number of vulnerabilities detected is lower. This mismatch occurs because the tool cannot find the correct version of the package, and consequently, the vulnerabilities associated with it.

Finally, the \texttt{OS} technique only affects Syft (regarding the number of CVEs), Scout, and Google Artifact Registry. This result is explained by the fact that Syft is unable to find the correct CVE information because it misses OS information, and the other tools do not search for package information without OS information. The tools that are resilient against this technique either find OS information in other layers or by other means (e.g., metadata information of the container).

In contrast, ORCA is not affected by any of the single-obscuration techniques.

\paragraph{Multiple obscuration}
We then analyzed the effect of multiple techniques on the same container. Syft, in the non-default configuration, is the only tool not affected by a combination of two or more obscuration techniques (unless one of them is the URL technique). OS and PKG techniques combined reduce the number of vulnerabilities discovered to zero in Google Artifact Analysis and Docker Scout, while in Syft (with default configuration), the number of vulnerabilities goes from 625 to 3. In the case of 3 concurrent obscurations, 5 of the 6 tools analyzed report zero vulnerabilities. We noticed that Google Artifact Registry is unable to produce results when the \texttt{OS} and \texttt{PKG} techniques are applied together. Amazon does not produce results when the OS package manager is not available. 

Overall, the effect of obscuration appears to be significant; we were able to reduce the number of vulnerabilities to 0 in every tool but Syft (which still has a $\approx$ 97\% reduction in the number of vulnerabilities). 

\paragraph{Impact of multi-stage builds}
The only technique effective against all tools, including our own, is observed when a container uses a multi-stage build following other tactics. In such instances, intermediate artifacts generated in earlier layers are systematically removed and cannot be recovered.

\subsection{Obscuration in popular containers (RQ3)}
\label{ssec:rq4}

This experiment investigates the existence of obscure containers in popular container registries, using the categories defined in Section~\ref{defining-container-image-obscuration}. We selected Python, Ruby, Node.js, PHP, and Go as target programming languages. Table \ref{tab:obscuration} summarizes the results, showing the number of containers with missing, modified, or deleted information, respectively, across the different obscuration types. We note that the \textit{Missing} column applies only to OS, OSPKG, and URL information, which are expected in a standard container filesystem, whereas language-specific dependencies or packages (DEP and PKG) appear only if the corresponding programming language is included.

More than 10\% of the containers present \texttt{OS} obscuration. Among them, the majority do not have OS information in any layer of the container, approximately one-third have modified OS information, and only three containers delete OS information in one of the layers. Almost 20\% of the containers download software directly from the Internet without using package managers.
\texttt{OSPKG} obscuration manifests in more than 50\% of the tested containers. Most of the instances of this misconfiguration modify the content of OS packages, which in many cases is the result of running OS updates. Approximately 10\% of containers do not have any OSPKG information, and a similar number delete one or more OS packages in the container layers. The results on missing OS and OSPKG information are particularly important as they highlight that tools such as Amazon Inspector and Google Artifact Registry would not be able to scan them. Similarly, results on \texttt{URL} obscuration show that over 30\% of the containers hide externally downloaded packages.

Finally, we found DEP and PKG obscuration on 3\% and 15\% of the containers, respectively. In such cases, it would be extremely complex to detect if a software dependency or package is missing because not all containers are supposed to have software dependencies. 

Another analysis consists of comparing the prevalence of obscuration across containers from different registries. Table \ref{tab:obscuration_registries} shows the number of containers with missing, modified, or deleted information, for different registries across various obscuration types. A first observation is that obscure containers exist in all container registries, from DockerHub to third-party registries. Among all datasets, DockerHub Bitnami seems to be the only one with no instances of by DEP or PKG obscuration. This happens because Bitnami uses tools to automatically compress the container images, removing the layer's history. However, the tool used by Bitnami (crane~\cite{Crane}) keeps the original Containerfile information as metadata. The analysis of Bitnami containers revealed that over 90\% of them download and install code from the Internet without using package managers. We also observed that obscure containers tend to be more present in DockerHub OSS and third-party registries like Quay.io and ECR, where we observed a significant number of deleted or modified Python package information, among others.

\begin{table}[!htb]
\centering
\small
\begin{tabular}{lrrr}
 \toprule
 \textbf{Obscuration type} & \textbf{Missing} & \textbf{Modified} & \textbf{Deleted} \\
 \cmidrule(lr){1-1} \cmidrule(lr){2-4}
 \textbf{OS} & 49 & 23 & 3 \\\cmidrule(lr){1-4}
 \textbf{OSPKG} & 55 & 364 & 58 \\\cmidrule(lr){1-4}
 \textbf{URL} & 187 & N/A & N/A \\  \cmidrule(lr){1-4}
 \textbf{DEP} & N/A & 15 & 3\\
 Python & N/A & 1 & 2\\
 Ruby & N/A & 2 & 1\\
 Node.js & N/A & 9 & 0\\
 PHP & N/A & 2 & 0\\
 Go & N/A & 1 & 0\\
 \cmidrule(lr){1-4}
 \textbf{PKG} & N/A & 69 & 23\\
 Python & N/A & 48 & 19\\
 Ruby & N/A & 5 & 1\\
 Node.js & N/A & 12 & 2\\
 PHP& N/A & 1 & 0\\
 Go & N/A & 3 & 1\\
 \bottomrule
\end{tabular}
\caption{Number of containers respectively with missing, modified, or deleted information across various obscuration types.}
\label{tab:obscuration}
\end{table}
\vspace{-2em}
\begin{table}[!htb]
\centering
\small
\begin{tabular}{lrrrrrr}
 \toprule
 \textbf{Obscuration type} & \textbf{DO} & \textbf{DB} & \textbf{DV} & \textbf{DOSS} & \textbf{Q} & \textbf{E} \\
 \cmidrule(lr){1-1} \cmidrule(lr){2-7}
 \textbf{OS} & 4 & 3 & 20 & 19 & 20 & 6 \\  \cmidrule(lr){1-7}
 \textbf{OSPKG} & 93 & 3 & 91 & 98 & 84 & 50 \\  \cmidrule(lr){1-7}
 \textbf{URL} & 27 & 94 & 13 & 39 & 9 & 5 \\  \cmidrule(lr){1-7}
 \textbf{DEP} & 0 & 0 & 0 & 7 & 6 & 7\\
 Python & 0 & 0 & 0 & 2 & 0 & 1 \\ 
 Ruby & 0 & 0 & 0 & 1 & 0 & 2 \\ 
 Node.js & 0 & 0 & 0 & 2 & 2 &  5\\ 
 PHP & 0 & 0 & 0 & 1 & 0 & 1\\
 Go & 0 & 0 & 0 & 0 & 0 & 1\\ 
 \cmidrule(lr){1-7}
\textbf{PKG} & 6 & 0 & 4 & 18 & 22 & 38 \\
 Python & 3 & 0 & 2 & 9 & 11 & 24 \\ 
 Ruby & 1 & 0 & 0 & 1 & 1 & 2\\ 
 Node.js & 2 & 0 & 1 & 3 & 4 & 2\\
 PHP & 0 & 0 & 0 & 0 & 0 & 1\\ 
 Go & 0 & 0 & 2 & 0 & 1 & 0\\ 
 \bottomrule
\end{tabular}
\caption{Number of containers with missing/modified/deleted information for six different registries across various obscuration types. DO: Docker Official. DB: Docker Bitnami. DV: Docker Verified. DOSS: Docker OSS. Q: Quay.io. E: ECR.}
\label{tab:obscuration_registries}
\end{table}

\subsection{Coverage and Performance}
This experiment evaluates the impact of our obscuration-resilient methodology on enhancing file coverage accuracy. For that, we used the same dataset as in Section~\ref{ssec:rq4}. We produced an SBOM of each container using the OSS tools (performing the same analysis using cloud tools was impractical) and ORCA.

\begin{figure}[h]
    \centering
    \begin{subfigure}[b]{0.48\textwidth}
        \centering
        \includegraphics[width=\textwidth]{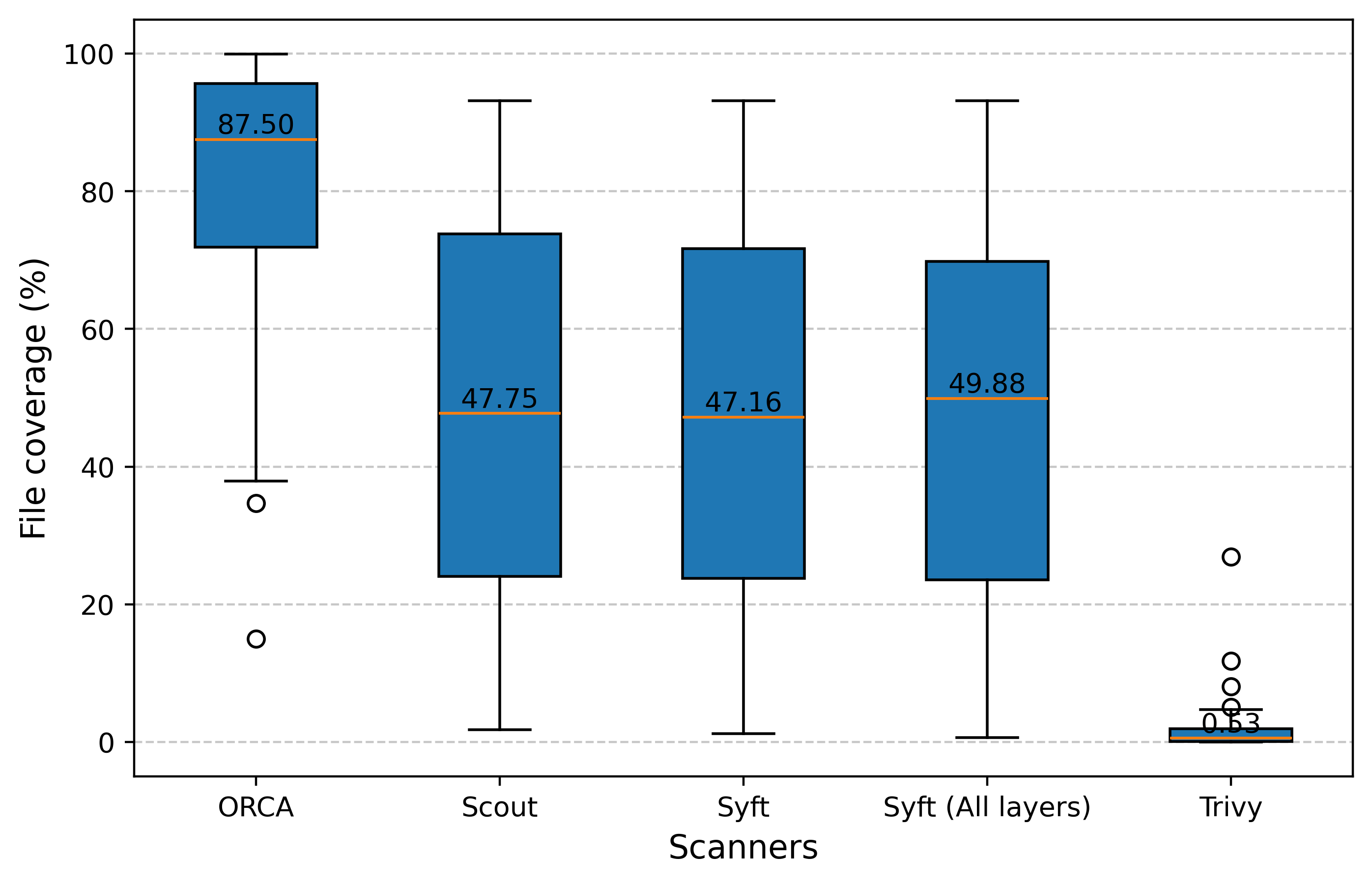}
        \caption{Box plot of file coverage per scanner. The red line indicates the median value.}
        \label{fig:file_coverage_boxplot}
    \end{subfigure}
    \hfill
    \begin{subfigure}[b]{0.48\textwidth}
        \centering
        \includegraphics[width=\textwidth]{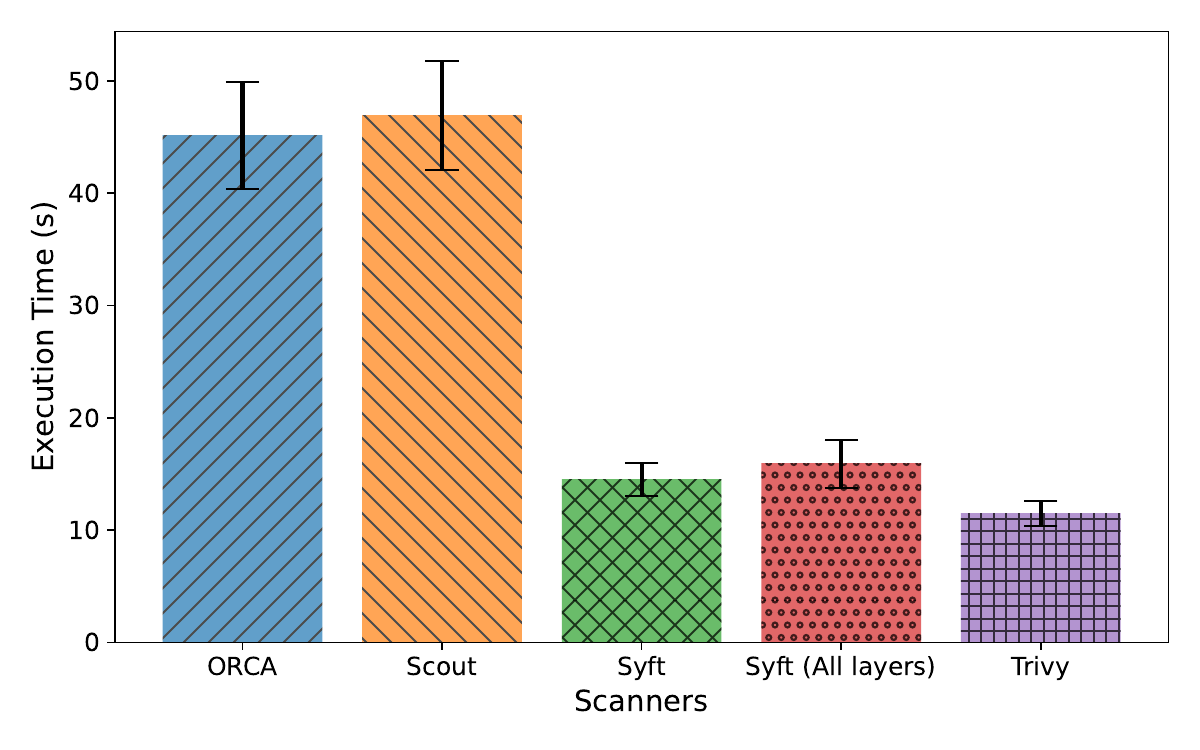}
        \caption{Average execution time of ORCA and other state-of-the-art tools. The black lines indicate the standard deviation.}
        \label{fig:execution_time}
    \end{subfigure}
    \caption{(a) File coverage per scanner and (b) average execution time of ORCA and other state-of-the-art tools.}
    \label{fig:coverage_and_time}
\end{figure}

As illustrated in Figure~\ref{fig:file_coverage_boxplot}, ORCA achieves a median file coverage of 87.5\%. This is possible because it analyzes files with package content and at the same time keeps track of the files related to each package. ORCA's result are $\approx$40\% higher than Scout and Syft and far exceeds Trivy's median file coverage of 0.53\%. The reason is that Scout and Syft only report the files installed by the package manager (dpkg), while Trivy only includes files that index packages.

Performance measurements are needed to assess the compatibility of ORCA with modern CI/CD pipelines. Therefore, we measured the execution time of ORCA and the other scanners (see Table~\ref{tab:scanners}) on the six datasets presented in Table~\ref{tab:datasets}. The average execution times are shown in Figure~\ref{fig:execution_time}. Our analysis reveals that ORCA took, on average, approximately 45 seconds to complete a scan, which is one second faster than Docker Scout and about 30 seconds slower than the other tools. The standard deviation is also limited (<10 seconds), indicating that ORCA scales well with the size of the container image. Considering ORCA's enhancements in resilience and filesystem coverage, the resulting execution time is suitable for typical CI/CD workflows.

%% file: related.tex
\section{Related work}
\label{sec:rw}

This section discusses recent works on Software Composition Analysis for containers and threats to the software supply chain.

\subsection{SCA for container images}

Recent research produced a large amount of literature on SBOMs, which can be divided into three main categories: SBOM generation, comparative analysis of SBOM tools, and identification of technical challenges. For SBOM generation tools, we refer the reader to Table~\ref{tab:scanners} that provides an overview of available tools for scanning container images. Studies such as~\cite{Torres23, Javed21, Yu24, o2023impacts} do not propose novel SCA methodologies but instead conduct qualitative analysis of existing SBOM generation tools for containers, comparing the tools' results across various datasets and evaluating their performance.
The work of Kawaguchi \textit{et al.}~\cite{Kawaguchi24} is the first to discuss the issues of external URL not being recognized by SCA tools. However, they do not discuss other causes of obscure containers.
Doan \textit{et al.}~\cite{CMC:2022:Doan} also noticed the possibility for containers to download packages from the Internet, but they focus on identifying so-called Potentially Vulnerable Files within container images, aiming to prioritize vulnerability detection over exhaustively finding all packages. 
These works highlight challenges, particularly in measuring and assuring two key metrics of SBOM generators: 1) \textbf{completeness}, which refers to how well the SBOM captures all components and dependencies, including both direct and transitive dependencies, and 2) \textbf{accuracy}, which assesses the correctness of the information provided in the SBOM, such as precise component names, versions, and metadata. Moreover, the authors in~\cite{Torres23} and~\cite{Halbritter24} assess the adherence of SBOM tools to the National Telecommunications and Information Administration (NTIA)'s minimum required elements~\cite{ntia2019sbom}, respectively for the SPDX and CycloneDX standards.

Given the prevalence of vulnerabilities in containers, several Docker Image Vulnerability Scanners have been developed~\cite{trivy,clair,docker_scout,snyk,Tern} to inspect containers and detect vulnerable components. These tools are relevant to SBOM generators, and some can detect vulnerabilities and generate SBOMs. Other works~\cite{ESCS:2020:Wist,ARXIV:2015:Bui,ACM:1027:Shu,Zhao21} discuss vulnerabilities of container images caused by third-party components, highlighting the importance of improving SCA techniques to reduce the risk of compromised containers.

\subsection{Attacks on the software supply chain}

Every link in the software supply chain can be targeted by malicious actors or weakened by misconfigurations and a lack of best practices. Ladisa \textit{et al.}~\cite{10179304} built an inventory of the main attacks and threats to the software supply chain. They identified 107 attack vectors and provided guidelines for both developers and third-party operators. Their primary focus is on malicious actors that either try to subvert legitimate packages --- by e.g., injecting malicious payloads in the application's source code --- or to create naming confusions with popular dependencies. Other works ~\cite{281396,10179471} have targeted the security of the build systems (CI/CD platforms), which should guarantee an isolated and trusted environment for building artifacts. The authors discovered numerous cases of over-permission and possible data leakage. 
In contrast, our work focuses on attacks that exploit weaknesses in SCA tools.

%% file: discussion.tex
\section{Discussion}
\label{sec:discussions}

This section summarizes the key takeaways of our research, discusses our interactions with cloud providers, and gives insights into future work. 

\subsection{Key results and takeaways}
\paragraph{Obscure images}
The first result of our research is that all SCA tools for containers are unable to detect and scan, to varying degrees, obscure images. Such images are often not detected because of an incomplete analysis process that does not take into account all of the layers, files, and externally downloaded software. 

\paragraph{Scanning vs Ignoring obscure images}
We observed that Amazon Inspector and Google Artifact Registry do not analyze containers that lack an operating system or a package manager, respectively. This appears to reflect a deliberate choice by the providers to disregard images that do not contain certain expected files. However, we argue that excluding such images is problematic for several reasons. First, non-scanned images are difficult to distinguish from genuinely non-vulnerable ones, and may therefore be incorrectly perceived as non-vulnerable in the UI. Second, developers may still rely on obscure images, for example, because they trust the correctness of their Containerfiles. Finally, containers without an operating system or package manager are not necessarily obscure by default, as images may legitimately consist of a single binary without targeting a specific operating system.

\paragraph{Limitations}
Several factors may affect the accuracy and validity of our results. First, we do not parse binary files, which may cause us to miss package information embedded in ELF/PE symbols. Second, all images used to generate the obscured containers are based on the same stack (a Debian base image with Python packages). Third, our methodology is effective against obscure containers only if the entire build process occurs inside the container. For example, we cannot detect \texttt{LINK} or \texttt{ALIAS} techniques if they are created outside the Containerfile. This limits the generalizability of our results, as the evaluation may not fully capture resilience across different distributions, ecosystems, or polyglot container environments. In addition, we do not analyze baseline differences among tools, which prevents us from directly comparing them. Finally, our tool ORCA was optimized for the operating systems and package managers present in our dataset.

\paragraph{Comparison of SCA tools}
The current literature on SCA tends to measure the performance of container scanning tools based on the number of CVEs they detect. In this paper, we demonstrated that the number of vulnerabilities is not a correct measure of the completeness of a scan. On the contrary, it may give a false sense of security because vulnerabilities may be duplicated or not applicable to a specific context (e.g., development dependency is not installed or the vulnerability relates to a different architecture). Similarly, the number of detected packages alone is not sufficient to demonstrate the performance of a tool. In fact, some tools display Windows packages on Linux containers or have multiple entries for the same package. In this work, we demonstrated that file-system coverage is a better metric because it indicates how many files cannot be identified as packages or package-related content.

\subsection{Mitigations and Remediations}

The main mitigation against obscure containers is ensuring transparency in the container image contents (i.e., make them easy to scan). In practice, this means guaranteeing that the files used to index packages are included in the final container artifact. We have identified five actionable remediations developers can adopt to avoid building obscure images:

\paragraph{Reduce image size responsibly:} Reducing space in a container image should not involve manually deleting operating system files (e.g., \texttt{rm -rf /etc/*}). This practice can remove critical metadata needed for vulnerability scanning and software auditing. Instead, use tools like \textit{docker-slim}~\cite{Slim} or minimal base images to reduce image size safely.

\paragraph{Use of multi-stage builds:} When using multi-stage builds, we recommend copying dependency information files along with the final artifacts. These files are often discarded in the final stage, but are essential for tracking software components. Retaining them enables security tools to perform accurate analysis of all installed packages.

\paragraph{Correct software installation} Prefer installing software using the operating system's package manager. This ensures that proper metadata and dependency information are preserved, enabling better security analysis. Manual or source-based installations may bypass package tracking, reducing visibility into potential vulnerabilities.

\paragraph{Transparency of compiled software} When software must be compiled from source, ensure that package-identifiable metadata is preserved (e.g., \texttt{.git/} folder in the case of public repositories). This allows for the detection of vulnerable components and supports better traceability of build-time configurations.

\paragraph{Verify supported ecosystems} Each Software Composition Analysis (SCA) tool supports different package managers and programming languages. Ensuring compatibility with the contents of the container image helps avoid false negatives during analysis.

%% file: conclusion.tex
\section{Conclusion and Future Work}
\label{sec:ccl}
This paper explores novel attacks on the software supply chain, specifically targeting software containers and SBOM (Software Bill of Materials) documents. Through extensive evaluation of both open-source and proprietary cloud tools, we demonstrate their lack of resilience against such attacks. We introduce new methodologies to uncover obscure container images and mitigate their associated attack surfaces. A comprehensive analysis of 600 popular containers enabled us to identify hundreds of such obscure images. We show that our methodology improves filesystem coverage by up to 600\% compared to state-of-the-art tools. We responsibly disclosed our findings to the maintainers of various SCA tools and engaged in constructive discussions. In the future, we anticipate broader adoption of coverage-based SCA mechanisms. As part of future work, we plan to extend our study to include malicious obfuscation and analysis of entry-point scripts. Another potential area of study is understanding the differences in the tools' baseline results regarding packages and vulnerabilities.

%% file: appendix.tex
\section{Package obscuration location}
\label{app:location}
This appendix contains detailed information about the types and patterns of files used to detect obscuration in our work. They are presented in Table~\ref{tab:obscuration:patterns}. 

\begin{table}[b]
\centering
\begin{tabular}{l c l}
 \toprule
 \textbf{ID} & \textbf{Type} &  \textbf{Pattern} \\
  \cmidrule(lr){1-3} 
 
OS  & Any & \texttt{os-release}, \texttt{etc-release} \\
    &     &  \texttt{debian\_version} \\  \cmidrule(lr){1-3}
      
OSPKG & DPKG & \texttt{dpkg\slash status},\texttt{var\slash lib\slash dpkg} \\  
      & RPM & \texttt{rpm\slash Packages}, \texttt{rpmdb.sqlite} \\
      &    & \texttt{var\slash lib\slash yum}, \texttt{var\slash cache\slash yum} \\
      &    & \texttt{yum.repos.d} \\ 
 
     & APK & \texttt{apk\slash db\slash installed},\texttt{apk\slash world} \\   \cmidrule(lr){1-3}
 
DEP & Python & \texttt{Pipfile}, \texttt{requirements.txt} \\
    & Ruby   & \texttt{.gemspec} \\
    & Node.js& \texttt{package.json}, \texttt{package-lock.json} \\
    &        & \texttt{yarn.lock} \\
    & PHP    & \texttt{composer.json}, \texttt{composer.lock} \\
    & Go     & \texttt{go.sum}, \texttt{go.mod} \\   \cmidrule(lr){1-3}

 PKG & Python & \texttt{dist-info\slash }, \texttt{egg-info\slash } \\
     &        & \texttt{site-packages\slash }, \texttt{dist-packages\slash } \\
     & Ruby   & \texttt{gems\slash} \\
     & Node.js& \texttt{node\_modules\slash} \\
     & PHP    & \texttt{\slash vendor\slash} \\
     & Go     & \texttt{\slash go\slash} \\\bottomrule
     
\end{tabular}
\caption{Patterns used for obscuration detection. For each file in a container, if the file matches one of the patterns and its history shows modifications, then is considered  obscure}
\label{tab:obscuration:patterns}
\end{table}

\section{Example changes in Containerfiles}
\label{ssec:example_obscure}

The following illustrates a few examples of changes in a Containerfile that ease the scan of the image. The first example in Figure~\ref{fig:obscurejs} shows two different Containerfiles for a JavaScript (React) application. Both images uses multi-stage builds. The obscure version (Fig.~\ref{fig:obscure:obscure}) copies to the final container layer only the build artifact which contains an \texttt{index.html} and a minified JavaScript file. This image is hard to scan because there is no direct information on the dependencies used to build and package the application. The other version of the same Containerfile (Fig.~\ref{fig:obscure:fixed}) includes the package metadata, allowing SCA tools to easy find dependencies and possible CVEs. Figure~\ref{fig:obscurepkg} shows instead two different ways to install a package: from source and using the package manager. The first version is impossible to scan using state-of-the-art SCA tools, while the second version is instead easy to scan.
\begin{figure}[htb!]
     \centering
    \subcaptionbox{Obscure Dockerfile\label{fig:obscure:obscure}}[.8675\linewidth]{%
\begin{lstlisting}[style=cyaml]
# Stage 1 - Build app
FROM node:18 AS builder
WORKDIR /app
COPY . .
RUN npm install && npm run build

# Stage 2 - Serve with nginx
FROM nginx:alpine
COPY --from=builder /app/build
 /usr/share/nginx/html
\end{lstlisting}
    }
    \vspace{1em} 

    \subcaptionbox{Improved Dockerfile\label{fig:obscure:fixed}}[.8675\linewidth]{%
\begin{lstlisting}[style=cyaml]
# Stage 1 - Build app
FROM node:18 AS builder
WORKDIR /app
COPY . .
RUN npm install && npm run build

# Stage 2 - Serve with nginx
FROM nginx:alpine
COPY --from=builder /app/build 
  /usr/share/nginx/html
COPY --from=builder /app/package*.json /app
\end{lstlisting}
    }

    \caption{%
        \subref{fig:obscure:obscure}) An example of an obscure JavaScript container build. 
        \subref{fig:obscure:fixed}) Same container made more transparent and easier to scan.
    }
    \label{fig:obscurejs}

\end{figure}

\begin{figure}[htb!]
     \centering
    \subcaptionbox{Obscure Dockerfile\label{fig:obscurepkg:obscure}}[.8675\linewidth]{%
\begin{lstlisting}[style=cyaml]
FROM debian:bullseye AS builder
ENV CURL_VERSION=8.7.1
RUN curl -LO https://curl.se/download/curl-
    ${CURL_VERSION}.tar.gz && \
    tar -xzf curl-${CURL_VERSION}.tar.gz && \
    cd curl-${CURL_VERSION} && \
    ./configure --with-ssl && \
    make -j$(nproc) && make install && \
    cd .. && rm -rf curl-${CURL_VERSION}*
\end{lstlisting}
    }
    \vspace{1em} 

    \subcaptionbox{Improved Dockerfile\label{fig:obscurepkg:fixed}}[.8675\linewidth]{%
\begin{lstlisting}[style=cyaml]
# Stage 1 - Build
FROM debian:bullseye AS builder
RUN apt-get install -y curl

\end{lstlisting}
    }

    \caption{%
        \subref{fig:obscurepkg:obscure}) OS binary (OSPKG) package installed from source.
        \subref{fig:obscurepkg:fixed}) The same package installed via the package
         manager.
    }
    \label{fig:obscurepkg}

\end{figure}

%% file: main.bbl

\begin{thebibliography}{56}

%%% the Web address.

\ifx \showCODEN    \undefined \def \showCODEN     #1{\unskip}     \fi
\ifx \showDOI      \undefined \def \showDOI       #1{#1}\fi
\ifx \showISBNx    \undefined \def \showISBNx     #1{\unskip}     \fi
\ifx \showISBNxiii \undefined \def \showISBNxiii  #1{\unskip}     \fi
\ifx \showISSN     \undefined \def \showISSN      #1{\unskip}     \fi
\ifx \showLCCN     \undefined \def \showLCCN      #1{\unskip}     \fi
\ifx \shownote     \undefined \def \shownote      #1{#1}          \fi
\ifx \showarticletitle \undefined \def \showarticletitle #1{#1}   \fi
\ifx \showURL      \undefined \def \showURL       {\relax}        \fi
% invisible to TeX
\providecommand\bibfield[2]{#2}
\providecommand\bibinfo[2]{#2}
\providecommand\natexlab[1]{#1}
\providecommand\showeprint[2][]{arXiv:#2}

\bibitem[{Amazon Web Services, Inc.}(2024a)]%
        {AWSSbom}
\bibfield{author}{\bibinfo{person}{{Amazon Web Services, Inc.}}} \bibinfo{year}{2024}\natexlab{a}.
\newblock \bibinfo{title}{Amazon Inspector}.
\newblock \bibinfo{howpublished}{\url{https://aws.amazon.com/inspector/}}.
\newblock
\newblock
\shownote{Accessed: 2025-05-10}.


\bibitem[{Amazon Web Services, Inc.}(2024b)]%
        {ECR}
\bibfield{author}{\bibinfo{person}{{Amazon Web Services, Inc.}}} \bibinfo{year}{2024}\natexlab{b}.
\newblock \bibinfo{title}{Elastic {C}ontainer {R}egistry ({ECR})}.
\newblock \bibinfo{howpublished}{\url{https://aws.amazon.com/ecr/}}.
\newblock
\newblock
\shownote{Accessed: 2025-05-20}.


\bibitem[{Anchore, Inc.}(2024a)]%
        {grype}
\bibfield{author}{\bibinfo{person}{{Anchore, Inc.}}} \bibinfo{year}{2024}\natexlab{a}.
\newblock \bibinfo{title}{Grype: A {V}ulnerability {S}canner for {C}ontainer {I}mages and {F}ilesystems}.
\newblock \bibinfo{howpublished}{\url{https://github.com/anchore/grype}}.
\newblock
\newblock
\shownote{Accessed: 2025-05-22}.


\bibitem[{Anchore, Inc.}(2024b)]%
        {syft}
\bibfield{author}{\bibinfo{person}{{Anchore, Inc.}}} \bibinfo{year}{2024}\natexlab{b}.
\newblock \bibinfo{title}{{Syft: CLI tool and library for generating a Software Bill of Materials from container images and filesystems}}.
\newblock \bibinfo{howpublished}{\url{https://github.com/anchore/syft}}.
\newblock
\newblock
\shownote{Accessed: 2024-11-20}.


\bibitem[Brad et~al\mbox{.}(2023)]%
        {maliciouscompliance}
\bibfield{author}{\bibinfo{person}{Geesaman Brad}, \bibinfo{person}{Coldwater Ian}, \bibinfo{person}{McCune Rory}, {and} \bibinfo{person}{Cooley Duffie}.} \bibinfo{year}{2023}\natexlab{}.
\newblock \showarticletitle{Malicious Compliance: Reflections on Trusting Container Scanners}. In \bibinfo{booktitle}{\emph{{KubeCon Europe 2023}}}. \bibinfo{publisher}{Cloud Native Computing Foundation (CNCF)}.
\newblock


\bibitem[Brady et~al\mbox{.}(2020)]%
        {CCWC:2020:Brady}
\bibfield{author}{\bibinfo{person}{Kelly Brady}, \bibinfo{person}{Seung Moon}, \bibinfo{person}{Tuan Nguyen}, {and} \bibinfo{person}{Joel Coffman}.} \bibinfo{year}{2020}\natexlab{}.
\newblock \showarticletitle{Docker Container Security in Cloud Computing}. In \bibinfo{booktitle}{\emph{2020 10th Annual Computing and Communication Workshop and Conference (CCWC)}}. \bibinfo{pages}{0975--0980}.
\newblock
\urldef\tempurl%
\url{https://doi.org/10.1109/CCWC47524.2020.9031195}
\showDOI{\tempurl}


\bibitem[Brandon and Isaac(2024)]%
        {google_sbom}
\bibfield{author}{\bibinfo{person}{Lum Brandon} {and} \bibinfo{person}{Hepworth Isaac}.} \bibinfo{year}{2024}\natexlab{}.
\newblock \showarticletitle{Lessons Learned from Generating 100m SBOMs Google's Approach to SBOM Compliance}. In \bibinfo{booktitle}{\emph{{KubeCon Europe 2024}}}. \bibinfo{publisher}{Cloud Native Computing Foundation (CNCF)}.
\newblock


\bibitem[Bui(2015)]%
        {ARXIV:2015:Bui}
\bibfield{author}{\bibinfo{person}{Thanh Bui}.} \bibinfo{year}{2015}\natexlab{}.
\newblock \bibinfo{title}{Analysis of Docker Security}.
\newblock
\newblock
\showeprint[arxiv]{1501.02967}~[cs.CR]
\urldef\tempurl%
\url{https://arxiv.org/abs/1501.02967}
\showURL{%
\tempurl}


\bibitem[Doan and Jung(2022)]%
        {CMC:2022:Doan}
\bibfield{author}{\bibinfo{person}{Phuc Doan} {and} \bibinfo{person}{Souhwan Jung}.} \bibinfo{year}{2022}\natexlab{}.
\newblock \showarticletitle{{{DAVS}}: {{Dockerfile Analysis}} for {{Container Image Vulnerability Scanning}}}.
\newblock \bibinfo{journal}{\emph{Computers, Materials \& Continua}}  \bibinfo{volume}{72} (\bibinfo{date}{Jan.} \bibinfo{year}{2022}), \bibinfo{pages}{1699--1711}.
\newblock
\urldef\tempurl%
\url{https://doi.org/10.32604/cmc.2022.025096}
\showDOI{\tempurl}


\bibitem[{Docker, Inc.}(2024a)]%
        {docker_scout}
\bibfield{author}{\bibinfo{person}{{Docker, Inc.}}} \bibinfo{year}{2024}\natexlab{a}.
\newblock \bibinfo{title}{Docker {Scout}}.
\newblock \bibinfo{howpublished}{\url{https://docs.docker.com/scout/}}.
\newblock
\newblock
\shownote{Accessed: 2025-05-22}.


\bibitem[{Docker, Inc.}(2024b)]%
        {DockerfileReference}
\bibfield{author}{\bibinfo{person}{{Docker, Inc.}}} \bibinfo{year}{2024}\natexlab{b}.
\newblock \bibinfo{title}{Dockerfile}.
\newblock \bibinfo{howpublished}{\url{https://docs.docker.com/reference/dockerfile/}}.
\newblock
\newblock
\shownote{Accessed: 2025-05-20}.


\bibitem[{Docker, Inc.}(2024c)]%
        {DockerHub}
\bibfield{author}{\bibinfo{person}{{Docker, Inc.}}} \bibinfo{year}{2024}\natexlab{c}.
\newblock \bibinfo{title}{DockerHub}.
\newblock \bibinfo{howpublished}{\url{https://hub.docker.com/}}.
\newblock
\newblock
\shownote{Accessed: 2025-05-20}.


\bibitem[Duan et~al\mbox{.}(2020)]%
        {duan2020towards}
\bibfield{author}{\bibinfo{person}{Ruian Duan}, \bibinfo{person}{Omar Alrawi}, \bibinfo{person}{Ranjita~Pai Kasturi}, \bibinfo{person}{Ryan Elder}, \bibinfo{person}{Brendan Saltaformaggio}, {and} \bibinfo{person}{Wenke Lee}.} \bibinfo{year}{2020}\natexlab{}.
\newblock \showarticletitle{Towards measuring supply chain attacks on package managers for interpreted languages}.
\newblock \bibinfo{journal}{\emph{arXiv preprint arXiv:2002.01139}} (\bibinfo{year}{2020}).
\newblock


\bibitem[Enck and Williams(2022)]%
        {IEEESP:2022:ENCK}
\bibfield{author}{\bibinfo{person}{William Enck} {and} \bibinfo{person}{Laurie Williams}.} \bibinfo{year}{2022}\natexlab{}.
\newblock \showarticletitle{Top Five Challenges in Software Supply Chain Security: Observations From 30 Industry and Government Organizations}.
\newblock \bibinfo{journal}{\emph{{IEEE Security \& Privacy}}} \bibinfo{volume}{20}, \bibinfo{number}{2} (\bibinfo{year}{2022}), \bibinfo{pages}{96--100}.
\newblock
\urldef\tempurl%
\url{https://doi.org/10.1109/MSEC.2022.3142338}
\showDOI{\tempurl}


\bibitem[{European Union}(2024)]%
        {cyberrecilience}
\bibfield{author}{\bibinfo{person}{{European Union}}.} \bibinfo{year}{2024}\natexlab{}.
\newblock \bibinfo{title}{Regulation 2024/2847 (Cyber Resilience Act)}.
\newblock \bibinfo{howpublished}{\url{https://eur-lex.europa.eu/eli/reg/2024/2847}}.
\newblock
\newblock
\shownote{Accessed: 2025-05-20}.


\bibitem[Flauzac et~al\mbox{.}(2020)]%
        {flauzacReviewNativeContainer2020}
\bibfield{author}{\bibinfo{person}{Olivier Flauzac}, \bibinfo{person}{Fabien Mauhourat}, {and} \bibinfo{person}{Florent Nolot}.} \bibinfo{year}{2020}\natexlab{}.
\newblock \showarticletitle{A Review of Native Container Security for Running Applications}.
\newblock \bibinfo{journal}{\emph{Procedia Computer Science}}  \bibinfo{volume}{175} (\bibinfo{date}{Jan.} \bibinfo{year}{2020}), \bibinfo{pages}{157--164}.
\newblock
\showISSN{1877-0509}
\urldef\tempurl%
\url{https://doi.org/10.1016/j.procs.2020.07.025}
\showDOI{\tempurl}


\bibitem[Fourné et~al\mbox{.}(2023)]%
        {IEEESP:2023:Fourne}
\bibfield{author}{\bibinfo{person}{Marcel Fourné}, \bibinfo{person}{Dominik Wermke}, \bibinfo{person}{Sascha Fahl}, {and} \bibinfo{person}{Yasemin Acar}.} \bibinfo{year}{2023}\natexlab{}.
\newblock \showarticletitle{A Viewpoint on Human Factors in Software Supply Chain Security: A Research Agenda}.
\newblock \bibinfo{journal}{\emph{{IEEE Security \& Privacy}}} \bibinfo{volume}{21}, \bibinfo{number}{6} (\bibinfo{year}{2023}), \bibinfo{pages}{59--63}.
\newblock
\urldef\tempurl%
\url{https://doi.org/10.1109/MSEC.2023.3316569}
\showDOI{\tempurl}


\bibitem[{Google, Inc.}(2024a)]%
        {GCloud}
\bibfield{author}{\bibinfo{person}{{Google, Inc.}}} \bibinfo{year}{2024}\natexlab{a}.
\newblock \bibinfo{title}{Artifact {A}nalysis}.
\newblock \bibinfo{howpublished}{\url{https://cloud.google.com/artifact-analysis}}.
\newblock
\newblock
\shownote{Accessed: 2025-05-10}.


\bibitem[{Google, Inc.}(2024b)]%
        {Crane}
\bibfield{author}{\bibinfo{person}{{Google, Inc.}}} \bibinfo{year}{2024}\natexlab{b}.
\newblock \bibinfo{title}{Crane}.
\newblock \bibinfo{howpublished}{\url{https://github.com/google/go-containerregistry}}.
\newblock
\newblock
\shownote{Accessed: 2025-05-20}.


\bibitem[Gu et~al\mbox{.}(2023)]%
        {10179471}
\bibfield{author}{\bibinfo{person}{Yacong Gu}, \bibinfo{person}{Lingyun Ying}, \bibinfo{person}{Huajun Chai}, \bibinfo{person}{Chu Qiao}, \bibinfo{person}{Haixin Duan}, {and} \bibinfo{person}{Xing Gao}.} \bibinfo{year}{2023}\natexlab{}.
\newblock \showarticletitle{Continuous Intrusion: Characterizing the Security of Continuous Integration Services}. In \bibinfo{booktitle}{\emph{2023 IEEE Symposium on Security and Privacy (SP)}}. \bibinfo{publisher}{IEEE}, \bibinfo{address}{3 Park Avenue, 17th Floor, New York, NY 10016-5997, USA}, \bibinfo{pages}{1561--1577}.
\newblock
\urldef\tempurl%
\url{https://doi.org/10.1109/SP46215.2023.10179471}
\showDOI{\tempurl}


\bibitem[Halbritter and Merli(2024)]%
        {Halbritter24}
\bibfield{author}{\bibinfo{person}{Andreas Halbritter} {and} \bibinfo{person}{Dominik Merli}.} \bibinfo{year}{2024}\natexlab{}.
\newblock \showarticletitle{Accuracy Evaluation of SBOM Tools for Web Applications and System-Level Software}. In \bibinfo{booktitle}{\emph{Proceedings of the 19th International Conference on Availability, Reliability and Security}} (Vienna, Austria) \emph{(\bibinfo{series}{ARES '24})}. \bibinfo{publisher}{Association for Computing Machinery}, \bibinfo{address}{New York, NY, USA}, Article \bibinfo{articleno}{55}, \bibinfo{numpages}{9}~pages.
\newblock
\showISBNx{9798400717185}
\urldef\tempurl%
\url{https://doi.org/10.1145/3664476.3670926}
\showDOI{\tempurl}


\bibitem[Herr(2021)]%
        {264122}
\bibfield{author}{\bibinfo{person}{Trey Herr}.} \bibinfo{year}{2021}\natexlab{}.
\newblock \showarticletitle{Breaking Trust {\textendash} Shades of Crisis Across an Insecure Software Supply Chain}. In \bibinfo{booktitle}{\emph{USENIX 2021}}. \bibinfo{publisher}{USENIX Association}, \bibinfo{address}{2560 Ninth Street, Suite 215, Berkeley, CA 94710, USA}.
\newblock


\bibitem[Imtiaz et~al\mbox{.}(2021)]%
        {imtiaz2021comparative}
\bibfield{author}{\bibinfo{person}{Nasif Imtiaz}, \bibinfo{person}{Seaver Thorn}, {and} \bibinfo{person}{Laurie Williams}.} \bibinfo{year}{2021}\natexlab{}.
\newblock \showarticletitle{A comparative study of vulnerability reporting by software composition analysis tools}. In \bibinfo{booktitle}{\emph{Proceedings of the 15th ACM/IEEE International Symposium on Empirical Software Engineering and Measurement (ESEM)}}. \bibinfo{pages}{1--11}.
\newblock


\bibitem[Inc(2024)]%
        {ScanImagesOS}
\bibfield{author}{\bibinfo{person}{Amazon Inc}.} \bibinfo{year}{2024}\natexlab{}.
\newblock \bibinfo{title}{Scan Images for {{OS}} and Programming Language Package Vulnerabilities in {{Amazon ECR}} - {{Amazon ECR}}}.
\newblock \bibinfo{howpublished}{\url{https://docs.aws.amazon.com/AmazonECR/latest/userguide/image-scanning-enhanced.html}}.
\newblock


\bibitem[Inc.(2024)]%
        {UnderstandingImageLayers0530}
\bibfield{author}{\bibinfo{person}{Docker Inc.}} \bibinfo{year}{2024}\natexlab{}.
\newblock \bibinfo{title}{Understanding the Image Layers}.
\newblock \bibinfo{howpublished}{\url{https://docs.docker.com/guides/docker-concepts/building-images/understanding-image-layers/}}.
\newblock
\newblock
\shownote{Accessed: 2025-05-17}.


\bibitem[{Intel, Inc.}(2024)]%
        {cve-bin}
\bibfield{author}{\bibinfo{person}{{Intel, Inc.}}} \bibinfo{year}{2024}\natexlab{}.
\newblock \bibinfo{title}{{cve-bin-tool}}.
\newblock \bibinfo{howpublished}{\url{https://github.com/intel/cve-bin-tool}}.
\newblock
\newblock
\shownote{Accessed: 2025-05-22}.


\bibitem[Javed and Toor(2021)]%
        {Javed21}
\bibfield{author}{\bibinfo{person}{Omar Javed} {and} \bibinfo{person}{Salman Toor}.} \bibinfo{year}{2021}\natexlab{}.
\newblock \showarticletitle{Understanding the Quality of Container Security Vulnerability Detection Tools}.
\newblock \bibinfo{journal}{\emph{arXiv preprint}} (\bibinfo{date}{01} \bibinfo{year}{2021}).
\newblock
\urldef\tempurl%
\url{https://doi.org/10.48550/arXiv.2101.03844}
\showDOI{\tempurl}


\bibitem[Kawaguchi et~al\mbox{.}(2024)]%
        {Kawaguchi24}
\bibfield{author}{\bibinfo{person}{Nobutaka Kawaguchi}, \bibinfo{person}{Charles Hart}, {and} \bibinfo{person}{Hiroki Uchiyama}.} \bibinfo{year}{2024}\natexlab{}.
\newblock \showarticletitle{Understanding the Effectiveness of SBOM Generation Tools for Manually Installed Packages in Docker Containers}.
\newblock \bibinfo{journal}{\emph{Journal of Internet Services and Information Security (JISIS)}} (\bibinfo{year}{2024}).
\newblock
\urldef\tempurl%
\url{https://doi.org/10.58346/JISIS.2024.I3.011}
\showDOI{\tempurl}


\bibitem[Koishybayev et~al\mbox{.}(2022)]%
        {281396}
\bibfield{author}{\bibinfo{person}{Igibek Koishybayev}, \bibinfo{person}{Aleksandr Nahapetyan}, \bibinfo{person}{Raima Zachariah}, \bibinfo{person}{Siddharth Muralee}, \bibinfo{person}{Bradley Reaves}, \bibinfo{person}{Alexandros Kapravelos}, {and} \bibinfo{person}{Aravind Machiry}.} \bibinfo{year}{2022}\natexlab{}.
\newblock \showarticletitle{Characterizing the Security of Github {CI} Workflows}. In \bibinfo{booktitle}{\emph{31st USENIX Security Symposium (USENIX Security 22)}}. \bibinfo{publisher}{USENIX Association}, \bibinfo{address}{Boston, MA}, \bibinfo{pages}{2747--2763}.
\newblock
\showISBNx{978-1-939133-31-1}
\urldef\tempurl%
\url{https://www.usenix.org/conference/usenixsecurity22/presentation/koishybayev}
\showURL{%
\tempurl}


\bibitem[Ladisa et~al\mbox{.}(2023)]%
        {10179304}
\bibfield{author}{\bibinfo{person}{Piergiorgio Ladisa}, \bibinfo{person}{Henrik Plate}, \bibinfo{person}{Matias Martinez}, {and} \bibinfo{person}{Olivier Barais}.} \bibinfo{year}{2023}\natexlab{}.
\newblock \showarticletitle{SoK: Taxonomy of Attacks on Open-Source Software Supply Chains}. In \bibinfo{booktitle}{\emph{2023 IEEE Symposium on Security and Privacy (SP)}}. \bibinfo{pages}{1509--1526}.
\newblock
\urldef\tempurl%
\url{https://doi.org/10.1109/SP46215.2023.10179304}
\showDOI{\tempurl}


\bibitem[Levy(2003)]%
        {1203227}
\bibfield{author}{\bibinfo{person}{E. Levy}.} \bibinfo{year}{2003}\natexlab{}.
\newblock \showarticletitle{Poisoning the software supply chain}.
\newblock \bibinfo{journal}{\emph{{IEEE Security \& Privacy}}} \bibinfo{volume}{1}, \bibinfo{number}{3} (\bibinfo{year}{2003}), \bibinfo{pages}{70--73}.
\newblock
\urldef\tempurl%
\url{https://doi.org/10.1109/MSECP.2003.1203227}
\showDOI{\tempurl}


\bibitem[Liu et~al\mbox{.}(2020)]%
        {ESORICS:2020:Liu}
\bibfield{author}{\bibinfo{person}{Peiyu Liu}, \bibinfo{person}{Shouling Ji}, \bibinfo{person}{Lirong Fu}, \bibinfo{person}{Kangjie Lu}, \bibinfo{person}{Xuhong Zhang}, \bibinfo{person}{Wei-Han Lee}, \bibinfo{person}{Tao Lu}, \bibinfo{person}{Wenzhi Chen}, {and} \bibinfo{person}{Raheem Beyah}.} \bibinfo{year}{2020}\natexlab{}.
\newblock \showarticletitle{Understanding the Security Risks of Docker Hub}. In \bibinfo{booktitle}{\emph{Computer Security – ESORICS 2020}}, \bibfield{editor}{\bibinfo{person}{Liqun Chen}, \bibinfo{person}{Ninghui Li}, \bibinfo{person}{Kaitai Liang}, {and} \bibinfo{person}{Steve Schneider}} (Eds.). \bibinfo{publisher}{Springer International Publishing}, \bibinfo{address}{Cham}, \bibinfo{pages}{257–276}.
\newblock
\showISBNx{978-3-030-58951-6}
\urldef\tempurl%
\url{https://doi.org/10.1007/978-3-030-58951-6_13}
\showDOI{\tempurl}


\bibitem[McCune({[n.\,d.]})]%
        {FunContainerImages}
\bibfield{author}{\bibinfo{person}{Rory McCune}.} \bibinfo{year}{[n.\,d.]}\natexlab{}.
\newblock \bibinfo{title}{Fun with Container Images - {{Bypassing}} Vulnerability Scanners}.
\newblock \bibinfo{howpublished}{https://raesene.github.io/blog/2023/04/22/Fun-with-container-images-Bypassing-vulnerability-scanners/}.
\newblock


\bibitem[Melara and Torres-Arias(2023)]%
        {IEEESP:2023:Melara}
\bibfield{author}{\bibinfo{person}{Marcela~S. Melara} {and} \bibinfo{person}{Santiago Torres-Arias}.} \bibinfo{year}{2023}\natexlab{}.
\newblock \showarticletitle{A Viewpoint on Software Supply Chain Security: Are We Getting Lost in Translation?}
\newblock \bibinfo{journal}{\emph{{IEEE Security \& Privacy}}} \bibinfo{volume}{21}, \bibinfo{number}{6} (\bibinfo{year}{2023}), \bibinfo{pages}{55--58}.
\newblock
\urldef\tempurl%
\url{https://doi.org/10.1109/MSEC.2023.3316568}
\showDOI{\tempurl}


\bibitem[{Microsoft, Inc.}(2024)]%
        {AzureSBOM}
\bibfield{author}{\bibinfo{person}{{Microsoft, Inc.}}} \bibinfo{year}{2024}\natexlab{}.
\newblock \bibinfo{title}{{SBOM Tool}}.
\newblock \bibinfo{howpublished}{\url{https://github.com/microsoft/sbom-tool}}.
\newblock
\newblock
\shownote{Accessed: 2025-05-10}.


\bibitem[{National Institute of Standards and Technology}(2024a)]%
        {NVD_CPE}
\bibfield{author}{\bibinfo{person}{{National Institute of Standards and Technology}}.} \bibinfo{year}{2024}\natexlab{a}.
\newblock \bibinfo{title}{{Common Platform Enumeration (CPE)}}.
\newblock \bibinfo{howpublished}{\url{https://nvd.nist.gov/products/cpe}}.
\newblock
\newblock
\shownote{Accessed: 2025-05-20}.


\bibitem[{National Institute of Standards and Technology}(2024b)]%
        {sbom}
\bibfield{author}{\bibinfo{person}{{National Institute of Standards and Technology}}.} \bibinfo{year}{2024}\natexlab{b}.
\newblock \bibinfo{title}{Executive Order 14028}.
\newblock \bibinfo{howpublished}{\url{https://www.nist.gov}}.
\newblock
\newblock
\shownote{Accessed: 2025-05-20}.


\bibitem[{National Institute of Standards and Technology}(2024c)]%
        {NVD}
\bibfield{author}{\bibinfo{person}{{National Institute of Standards and Technology}}.} \bibinfo{year}{2024}\natexlab{c}.
\newblock \bibinfo{title}{{National Vulnerability Database}}.
\newblock \bibinfo{howpublished}{\url{https://nvd.nist.gov/search}}.
\newblock
\newblock
\shownote{Accessed: 2025-05-20}.


\bibitem[{National Telecommunications and Information Administration (NTIA)}(2019)]%
        {ntia2019sbom}
\bibfield{author}{\bibinfo{person}{{National Telecommunications and Information Administration (NTIA)}}.} \bibinfo{year}{2019}\natexlab{}.
\newblock \bibinfo{title}{NTIA Software Bill of Materials (SBOM) Formats and Standards}.
\newblock
\newblock
\urldef\tempurl%
\url{https://www.ntia.gov/files/ntia/publications/ntia_sbom_formats_and_standards_whitepaper_-_version_20191025.pdf}
\showURL{%
\tempurl}
\newblock
\shownote{Version 20191025}.


\bibitem[O'Donoghue et~al\mbox{.}(2023)]%
        {o2023impacts}
\bibfield{author}{\bibinfo{person}{Eric O'Donoghue}, \bibinfo{person}{Brittany Boles}, \bibinfo{person}{Clemente Izurieta}, {and} \bibinfo{person}{Ann~Marie Reinhold}.} \bibinfo{year}{2023}\natexlab{}.
\newblock \showarticletitle{Impacts of software bill of materials (SBOM) generation on vulnerability detection}. In \bibinfo{booktitle}{\emph{Proceedings of the 2024 Workshop on Software Supply Chain Offensive Research and Ecosystem Defenses}}. \bibinfo{pages}{67--76}.
\newblock


\bibitem[{OWASP Foundation}(2021)]%
        {owaspTop10}
\bibfield{author}{\bibinfo{person}{{OWASP Foundation}}.} \bibinfo{year}{2021}\natexlab{}.
\newblock \bibinfo{title}{{OWASP Top Ten}}.
\newblock \bibinfo{howpublished}{\url{https://owasp.org/Top10}}.
\newblock
\urldef\tempurl%
\url{https://owasp.org/Top10}
\showURL{%
\tempurl}
\newblock
\shownote{Version 2021}.


\bibitem[Petersen et~al\mbox{.}(2015)]%
        {petersen2015guidelines}
\bibfield{author}{\bibinfo{person}{Kai Petersen}, \bibinfo{person}{Sairam Vakkalanka}, {and} \bibinfo{person}{Ludwik Kuzniarz}.} \bibinfo{year}{2015}\natexlab{}.
\newblock \showarticletitle{Guidelines for conducting systematic mapping studies in software engineering: An update}.
\newblock \bibinfo{journal}{\emph{Information and software technology}}  \bibinfo{volume}{64} (\bibinfo{year}{2015}), \bibinfo{pages}{1--18}.
\newblock


\bibitem[{Red Hat, Inc.}(2024)]%
        {QuayIO}
\bibfield{author}{\bibinfo{person}{{Red Hat, Inc.}}} \bibinfo{year}{2024}\natexlab{}.
\newblock \bibinfo{title}{Quay.io}.
\newblock \bibinfo{howpublished}{\url{https://quay.io/}}.
\newblock
\newblock
\shownote{Accessed: 2025-05-20}.


\bibitem[{RedHat, Inc.}(2024)]%
        {clair}
\bibfield{author}{\bibinfo{person}{{RedHat, Inc.}}} \bibinfo{year}{2024}\natexlab{}.
\newblock \bibinfo{title}{Clair: Vulnerability {S}tatic {A}nalysis for {C}ontainers}.
\newblock \bibinfo{howpublished}{\url{https://github.com/quay/clair}}.
\newblock
\newblock
\shownote{Accessed: 2025-05-22}.


\bibitem[Security(2024)]%
        {trivy}
\bibfield{author}{\bibinfo{person}{Aqua Security}.} \bibinfo{year}{2024}\natexlab{}.
\newblock \bibinfo{title}{Trivy: A Simple and Comprehensive Vulnerability Scanner for Containers and Other Artifacts}.
\newblock \bibinfo{howpublished}{\url{https://trivy.dev/}}.
\newblock
\newblock
\shownote{Accessed: 2025-05-22}.


\bibitem[Shu et~al\mbox{.}(2017)]%
        {ACM:1027:Shu}
\bibfield{author}{\bibinfo{person}{Rui Shu}, \bibinfo{person}{Xiaohui Gu}, {and} \bibinfo{person}{William Enck}.} \bibinfo{year}{2017}\natexlab{}.
\newblock \showarticletitle{A Study of Security Vulnerabilities on Docker Hub}. In \bibinfo{booktitle}{\emph{Proceedings of the Seventh ACM on Conference on Data and Application Security and Privacy}}. \bibinfo{publisher}{ACM}, \bibinfo{address}{Scottsdale Arizona USA}, \bibinfo{pages}{269–280}.
\newblock
\showISBNx{978-1-4503-4523-1}
\urldef\tempurl%
\url{https://doi.org/10.1145/3029806.3029832}
\showDOI{\tempurl}


\bibitem[{Slim.AI, Inc.}(2024)]%
        {Slim}
\bibfield{author}{\bibinfo{person}{{Slim.AI, Inc.}}} \bibinfo{year}{2024}\natexlab{}.
\newblock \bibinfo{title}{Slim: {Optimize and secure your containerized applications}}.
\newblock \bibinfo{howpublished}{\url{https://github.com/slimtoolkit/slim}}.
\newblock
\newblock
\shownote{Accessed: 2025-05-20}.


\bibitem[{Snyk}(2024)]%
        {snyk}
\bibfield{author}{\bibinfo{person}{{Snyk}}.} \bibinfo{year}{2024}\natexlab{}.
\newblock \bibinfo{title}{{Docker Security Scanning Guide}}.
\newblock \bibinfo{howpublished}{\url{https://snyk.io/articles/docker-security-scanning/}}.
\newblock
\newblock
\shownote{Accessed: 2024-12-19}.


\bibitem[Souppaya et~al\mbox{.}(2017)]%
        {NIST:2017:Souppaya}
\bibfield{author}{\bibinfo{person}{Murugiah Souppaya}, \bibinfo{person}{John Morello}, {and} \bibinfo{person}{Karen Scarfone}.} \bibinfo{year}{2017}\natexlab{}.
\newblock \bibinfo{booktitle}{\emph{Application container security guide}}.
\newblock \bibinfo{type}{{T}echnical {R}eport}. \bibinfo{institution}{National Institute of Standards and Technology}.
\newblock


\bibitem[Sultan et~al\mbox{.}(2019)]%
        {ContainerSecurityIssuesa}
\bibfield{author}{\bibinfo{person}{Sari Sultan}, \bibinfo{person}{Imtiaz Ahmad}, {and} \bibinfo{person}{Tassos Dimitriou}.} \bibinfo{year}{2019}\natexlab{}.
\newblock \bibinfo{title}{Container Security: Issues, Challenges, and the Road Ahead}.
\newblock , \bibinfo{numpages}{52976-52996}~pages.
\newblock
\urldef\tempurl%
\url{https://doi.org/10.1109/ACCESS.2019.2911732}
\showDOI{\tempurl}


\bibitem[{The Linux Foundation}(2024)]%
        {spdx}
\bibfield{author}{\bibinfo{person}{{The Linux Foundation}}.} \bibinfo{year}{2024}\natexlab{}.
\newblock \bibinfo{title}{{System Package Data Exchange (SPDX)}}.
\newblock \bibinfo{howpublished}{\url{https://spdx.dev/}}.
\newblock
\newblock
\shownote{Accessed: 2025-05-20}.


\bibitem[Tools(2024)]%
        {Tern}
\bibfield{author}{\bibinfo{person}{Tern Tools}.} \bibinfo{year}{2024}\natexlab{}.
\newblock \bibinfo{title}{Tern}.
\newblock \bibinfo{howpublished}{\url{https://github.com/tern-tools/tern}}.
\newblock
\newblock
\shownote{Accessed: 2025-05-20}.


\bibitem[Torres-Arias et~al\mbox{.}(2023)]%
        {Torres23}
\bibfield{author}{\bibinfo{person}{Santiago Torres-Arias}, \bibinfo{person}{Dan Geer}, {and} \bibinfo{person}{John~Speed Meyers}.} \bibinfo{year}{2023}\natexlab{}.
\newblock \showarticletitle{A Viewpoint on Knowing Software: Bill of Materials Quality When You See It}.
\newblock \bibinfo{journal}{\emph{IEEE Security \& Privacy}} \bibinfo{volume}{21}, \bibinfo{number}{6} (\bibinfo{year}{2023}), \bibinfo{pages}{50--54}.
\newblock
\urldef\tempurl%
\url{https://doi.org/10.1109/MSEC.2023.3315887}
\showDOI{\tempurl}


\bibitem[Wist et~al\mbox{.}(2021)]%
        {ESCS:2020:Wist}
\bibfield{author}{\bibinfo{person}{Katrine Wist}, \bibinfo{person}{Malene Helsem}, {and} \bibinfo{person}{Danilo Gligoroski}.} \bibinfo{year}{2021}\natexlab{}.
\newblock \showarticletitle{Vulnerability {{Analysis}} of 2500 {{Docker Hub Images}}}. In \bibinfo{booktitle}{\emph{Advances in {{Security}}, {{Networks}}, and {{Internet}} of {{Things}}}}, \bibfield{editor}{\bibinfo{person}{Kevin Daimi}, \bibinfo{person}{Hamid~R. Arabnia}, \bibinfo{person}{Leonidas Deligiannidis}, \bibinfo{person}{Min-Shiang Hwang}, {and} \bibinfo{person}{Fernando~G. Tinetti}} (Eds.). \bibinfo{publisher}{Springer International Publishing}, \bibinfo{address}{Cham}, \bibinfo{pages}{307--327}.
\newblock
\showISBNx{978-3-030-71017-0}
\urldef\tempurl%
\url{https://doi.org/10.1007/978-3-030-71017-0_22}
\showDOI{\tempurl}


\bibitem[Yu et~al\mbox{.}(2024)]%
        {Yu24}
\bibfield{author}{\bibinfo{person}{Sheng Yu}, \bibinfo{person}{Wei Song}, \bibinfo{person}{Xunchao Hu}, {and} \bibinfo{person}{Heng Yin}.} \bibinfo{year}{2024}\natexlab{}.
\newblock \showarticletitle{On the Correctness of Metadata-Based SBOM Generation: A Differential Analysis Approach}. In \bibinfo{booktitle}{\emph{2024 54th Annual IEEE/IFIP International Conference on Dependable Systems and Networks (DSN)}}. \bibinfo{pages}{29--36}.
\newblock
\urldef\tempurl%
\url{https://doi.org/10.1109/DSN58291.2024.00018}
\showDOI{\tempurl}


\bibitem[Zhao et~al\mbox{.}(2021)]%
        {Zhao21}
\bibfield{author}{\bibinfo{person}{Nannan Zhao}, \bibinfo{person}{Vasily Tarasov}, \bibinfo{person}{Hadeel Albahar}, \bibinfo{person}{Ali Anwar}, \bibinfo{person}{Lukas Rupprecht}, \bibinfo{person}{Dimitrios Skourtis}, \bibinfo{person}{Arnab~K. Paul}, \bibinfo{person}{Keren Chen}, {and} \bibinfo{person}{Ali~R. Butt}.} \bibinfo{year}{2021}\natexlab{}.
\newblock \showarticletitle{Large-Scale Analysis of Docker Images and Performance Implications for Container Storage Systems}.
\newblock \bibinfo{journal}{\emph{IEEE Transactions on Parallel Distributed Systems}} \bibinfo{volume}{32}, \bibinfo{number}{4} (\bibinfo{date}{April} \bibinfo{year}{2021}), \bibinfo{pages}{918–930}.
\newblock
\showISSN{1045-9219}
\urldef\tempurl%
\url{https://doi.org/10.1109/TPDS.2020.3034517}
\showDOI{\tempurl}


\end{thebibliography}
